%% file: main.tex
\documentclass[journal]{IEEEtran}
\usepackage{ieee_conference}
\input{tikz_prefs}

\input{acronyms}
\input{saving_space}

\begin{document}

\title{Generalized Framework for a Fair Comparison of Cellular and Cooperative Massive MIMO Systems}

\author{Leonard Paul Schulz\,\orcidlink{0000-0003-3385-2034}, \emph{Student Member, IEEE}, Stefan Schwarz\,\orcidlink{0000-0002-4065-2906}, \emph{Senior Member, IEEE}, \\and Gerhard Bauch\,\orcidlink{0000-0002-0050-2604}, \emph{Fellow, IEEE}%
\thanks{L. P. Schulz and G. Bauch are with the Institute of Communications, Hamburg University of Technology (TUHH), 21073 Hamburg, Germany (e-mail: leonard.schulz@tuhh.de; bauch@tuhh.de).}%
\thanks{S. Schwarz is with the Institute of Telecommunications, Technische Universit\"at Wien, 1040 Vienna, Austria (e-mail: stefan.schwarz@tuwien.ac.at). S. Schwarz has been funded by the Vienna Science and Technology Fund (WWTF) [Grant ID: 10.47379/ICT25005].}}

\maketitle
\glsdisablehyper
\input{abstract}
\vspace{-0.4em}
\glsresetall
\input{introduction}
\input{properties}
\input{system_model}
\input{signal_processing}
\input{eval_accuracy}
\input{results}
\input{conclusion}
\begingroup
\printbibliography
\endgroup
\end{document}

%% file: tikz_prefs.tex
\definecolor{tuhh_int_blue}{rgb}{0.00000, 0.27059, 0.54118}
\definecolor{tuhh_int_red}{rgb}{0.75294, 0.00000, 0.00000}
\definecolor{tuhh_int_black}{rgb}{0.00000, 0.00000, 0.00000}
\definecolor{tuhh_int_green_blue}{rgb}{0.09804, 0.47843, 0.51765}
\definecolor{tuhh_int_light_green}{rgb}{0.60784, 0.73333, 0.34902}
\definecolor{tuhh_int_orange}{rgb}{0.92549, 0.43137, 0.00000}
\definecolor{tuhh_int_dark_blue}{rgb}{0.00784, 0.14118, 0.25882}
\definecolor{tuhh_int_gray}{rgb}{0.64706, 0.64706, 0.64706}
\definecolor{tuhh_int_light_blue}{rgb}{0.34118, 0.74118, 0.76471}

\pgfplotscreateplotcyclelist{tuhh-int}{
    {tuhh_int_blue, mark=*, mark options={solid}},           
    {tuhh_int_red, mark=square*, mark options={solid}},   
    {tuhh_int_light_green, mark=triangle*, mark options={solid}}, 
    {tuhh_int_orange, mark=diamond*, mark options={solid}},     
    {tuhh_int_green_blue, mark=o, mark options={solid}},     
    {tuhh_int_black, mark=square, mark options={solid}},     
    {tuhh_int_gray, mark=triangle, mark options={solid}}, 
    {tuhh_int_dark_blue, mark=star, mark options={solid}},        
    {tuhh_int_light_blue, mark=+, mark options={solid}},     
}

\usetikzlibrary{arrows.meta,calc,scopes,spy}

\pgfplotsset{width=15cm,compat=1.18,cycle list name=tuhh-int}

\usepgfplotslibrary{external,fillbetween}
\tikzset{/tikz/external/optimize command away=\includepdf}
\AddToHook{env/algorithmic/begin}{\tikzexternaldisable}

%% file: acronyms.tex
\newacronym{siso}{SISO}{single-input single-output}
\newacronym{mimo}{MIMO}{multiple-input multiple-output}
\newacronym{mmimo}{mMIMO}{massive multiple-input multiple-output}
\newacronym{se}{SE}{spectral efficiency}

\newacronym{ue}{UE}{user equipment}
\newacronym{ap}{AP}{access point}
\newacronym{bs}{BS}{base station}
\newacronym{cpu}{CPU}{central processing unit}

\newacronym{dcc}{DCC}{dynamic cooperation clustering}

\newacronym{rf}{RF}{radio frequency}

\newacronym{csi}{CSI}{channel state information}
\newacronym{awgn}{AWGN}{additive white Gaussian noise}
\newacronym{sinr}{SINR}{signal-to-interference-plus-noise ratio}
\newacronym{rsrp}{RSRP}{reference signal received power}

\newacronym{tdd}{TDD}{time-division duplexing}

\newacronym{sp}{SP}{simplified partial}
\newacronym{lp}{LP}{local partial}
\newacronym{mmse}{MMSE}{minimum mean square error}
\newacronym{pmmse}{P-MMSE}{Partial MMSE}
\newacronym{psmmse}{PS-MMSE}{Partial Simplified MMSE}
\newacronym{lsfd}{LSFD}{large-scale fading decoding}
\newacronym{uatf}{UatF}{Use-and-then-Forget}
\newacronym{rzf}{RZF}{Regularized Zero-Forcing}
\newacronym{przf}{P-RZF}{Partial Regularized Zero-Forcing}
\newacronym{psrzf}{PS-RZF}{Partial Simplified Regularized Zero-Forcing}
\newacronym{lmmse}{L-MMSE}{local MMSE}
\newacronym{lpmmse}{LP-MMSE}{local partial MMSE}
\newacronym{mr}{MR}{maximum ratio}

\newacronym{pdf}{PDF}{probability density function}
\newacronym{cdf}{CDF}{cumulative distribution function}

\newacronym{rrh}{RRH}{remote radio head}
\newacronym{bbu}{BBU}{baseband unit}
\newacronym{ula}{ULA}{uniform linear array}
\newacronym{comp}{CoMP}{coordinated multipoint}
\newacronym{cran}{C-RAN}{cloud radio access network}
\newacronym{oran}{O-RAN}{Open Radio Access Network}
\newacronym{odu}{O-DU}{Open Distributed Unit}
\newacronym{oru}{O-RU}{Open Radio Unit}
\newacronym{cb}{CB}{conjugate beamforming}
\newacronym{ltea}{LTE-A}{Long Term Evolution Advanced}
\newacronym{3gpp}{3GPP}{3rd Generation Partnership Project}
\newacronym{das}{DAS}{Distributed Antenna System}
\newacronym{ppp}{PPP}{Poisson Point Process}

%% file: saving_space.tex
\usepackage{titlesec}

\AtEveryBibitem{
	\clearlist{language}
	\clearfield{doi}
	\clearfield{eprint}
	\clearfield{note}
	\clearfield{day}
	\clearfield{month}
	\clearfield{isbn}
	\ifentrytype{article}{
		\clearfield{issn}
	}{}
	\ifentrytype{inproceedings}{
		\clearfield{isbn}
		\clearfield{issn}
		\clearfield{url}
		\clearfield{urlyear}
	}{}
	\ifentrytype{thesis}{
		\clearfield{url}
		\clearfield{urlyear}
	}{}
}

%% file: abstract.tex
\begin{abstract}
Cooperative massive \gls{mimo} promises large gains over cellular deployments, but existing comparisons of different architectures often mix antenna distribution, inter-site coordination, and processing assumptions.
This paper introduces a graph-based framework for fair comparison of cellular, coordinated, and cell-free massive-\gls{mimo} systems.
We differentiate between two key properties, namely antenna distribution and inter-site cooperation, which yields seven representative system types.
We derive compatible uplink and downlink \gls{se} expressions, including an uplink bound for detectors with mixed instantaneous and statistical effective \gls{csi}, and adapt scalable user association and processing rules to all considered architectures.
We evaluate these systems using extensive numerical simulations and show that for a fair comparison much larger simulation areas (at least $2.5 \times \qty{2.5}{\kilo\meter\squared}$) than commonly used are required.
We introduce the \emph{relative capacity}, which measures how closely each architecture approaches centralized cell-free processing.
The results show that coordinated, phase-aligned beamforming across spatially distributed antennas is the main source of cooperation gains.
In dense deployments with few antennas per \gls{ap}, coordinated \gls{das} and hybrid cell-free architectures achieve much of the centralized cell-free performance while requiring substantially weaker midhaul assumptions.
\end{abstract}


%% file: introduction.tex
\section{Introduction}
Spatially distributing the antennas can dramatically improve the performance of a wireless network by providing macro-diversity and reducing the distance between the users and their serving antennas.
In combination with massive \gls{mimo} signal processing techniques, this approach has gained substantial interest as cell-free massive \gls{mimo}~\cite{ngoCellFreeMassiveMIMO2017,ngoUltradenseCellFreeMassive2024}.
The canonical cell-free architecture consists of a large number of geographically distributed \glspl{ap} that jointly serve all \glspl{ue} under the coordination of a single \gls{cpu} that the \glspl{ap} are connected to by ideal fronthaul links, essentially acting as one large distributed massive \gls{mimo} site.
While the foundational theory for such a system is by now well established \cite{bjornsonMakingCellFreeMassive2020,mirettiULDLDualityCellFree2024}, the practical implementation of cell-free massive \gls{mimo} remains challenging.
One fundamental requirement for a practical cell-free system is scalability, i.e., keeping the fronthaul and processing complexity manageable as the network grows in size \cite{interdonatoScalabilityAspectsCellFree2019,bjornsonScalableCellFreeMassive2020}.
Crucially, to make a system scalable, multiple \glspl{cpu} must be deployed.
If the \glspl{cpu} are interconnected via high-capacity and low-latency midhaul links they can be interpreted as a single virtual cloud \gls{cpu} which substantially simplifies the system model.
However, more recent works indicate that practical cell-free implementations, for example in \gls{oran}-type architectures, require an explicit interface for inter-\gls{cpu} coordination \cite{ranjbarCellFreeMMIMOSupport2022}.
If such an interface introduces limited capacity and non-negligible latency on the midhaul links, the system model has to account for the existence of multiple physical \glspl{cpu} explicitly, leading to new optimization challenges \cite{göttschFairnessSchedulingUserCentric2024,demirCellFreeMassiveMIMO2024}.

As the canonical cell-free architecture gets increasingly refined to account for practical implementation constraints, the boundary between advanced cellular coordination, such as \gls{comp} \cite{gesbertMultiCellMIMOCooperative2010}, and genuinely cell-free operation becomes less clear.
Our aim in this work is to establish a common basis for comparing cellular, coordinated, and cell-free massive \gls{mimo} systems under compatible modeling and processing assumptions by providing a unified framework that allows for different degrees of cooperation between sites.
In particular, we want to understand what kind of cooperation gives the largest gains in terms of achievable \gls{se} across different deployment scenarios.

\subsection{Related Work and Problem Statement}
Interest in cooperation in wireless networks dates back to the early 2000s, when cooperative relaying, virtual antenna arrays, and distributed transmission were studied as ways of exploiting spatial diversity beyond the limits of a single site (see e.g., \cite{fitzekCooperationWirelessNetworks2006} for a summary).
The theoretical potential of cooperative \gls{mimo} has also been investigated from an information-theoretic perspective under several idealized network models \cite{host-madsenCapacityBoundsCooperative2006,ozgurHierarchicalCooperationAchieves2007}.
In parallel, a substantial body of work considered cooperation between base stations in cellular networks, including coordinated beamforming, joint processing, relaying, and \gls{comp}-type architectures \cite{simeoneCooperativeWirelessCellular2012}.
Fundamental limits of such cooperation were studied in \cite{lozanoFundamentalLimitsCooperation2013,villacresFundamentalLimitsNoncoherent2025}.
These works provide important analytical insight, but they necessarily rely on tractable abstractions and therefore do not directly answer how different practical cooperation architectures should be compared under one common model.

The introduction of massive \gls{mimo} added a new perspective to cooperation because linear processing, \gls{tdd} operation, channel hardening arguments, and achievable \gls{se} bounds make large cooperative networks analytically and numerically more accessible \cite{marzettaNoncooperativeCellularWireless2010}, \cite{björnsonMassiveMIMONetworks2017}.
This led first to coordinated multi-cell massive-\gls{mimo} variants \cite{bjornsonOptimalResourceAllocation2013} and later to cell-free massive \gls{mimo} \cite{nayebiCellFreeMassiveMIMO2015}.
There is some comparison work between different versions of coordinated massive \gls{mimo} systems such as small-cell vs. cell-free \cite{ngoCellFreeMassiveMIMO2017}, centralized vs. distributed cell-free processing \cite{bjornsonMakingCellFreeMassive2020}, colocated vs. distributed antennas \cite{jiangUnifiedModelingPerformance2024}, transitional architectures between cells and full cell-free operation \cite{femeniasCellsFreedom6Gs2025}, multi-\gls{cpu} cooperation levels \cite{kimCPUCooperativePowerControl2024a}, and hybrid cell-free systems \cite{schulzHybridReceiveCombining2025}.

However, what is still missing is a complete overview comparing different cooperation types under a common framework that treats cellular, coordinated, and cell-free massive-\gls{mimo} systems with compatible assumptions.
Such a comparison should include both uplink and downlink operation, practical scalability constraints, explicit limits on coordination links, and effective signal processing with tight \gls{se} bounds, so that each architecture is evaluated close to its achievable performance and the comparison is not biased by architecture-dependent underestimation.

\subsection{Contributions}
In this work, we address the problem outlined above by introducing a generalized graph-based framework for cooperative massive-\gls{mimo} networks and using it to compare cellular, coordinated, and cell-free massive-\gls{mimo} systems under compatible modeling and processing assumptions.
In particular:
\begin{itemize}
      \item We rigorously define and identify seven representative system types that span colocated and distributed sites as well as different degrees of inter-site cooperation and categorize related work for each of these types.
      \item We derive a new uplink ergodic \gls{se} bound that remains tight across all considered system types, including cases where the detector has a mix of instantaneous and statistical effective \gls{csi}.
      In this way, we unify analyses that would otherwise need to switch between the classical instantaneous-\gls{csi} bound and the \gls{uatf} bound depending on the architecture \cite{bjornsonMakingCellFreeMassive2020}.
      \item We adapt high-performing scalable processing methods from the cell-free literature to the generalized multi-\gls{cpu} setting, including a joint initial-access, pilot-assignment, and association procedure \cite{bjornsonScalableCellFreeMassive2020}, hybrid \gls{mmse} combining \cite{schulzHybridReceiveCombining2025} with \gls{cpu}-level fusion in the uplink, and duality-motivated downlink precoding under per-\gls{ap} power constraints.
      \item We develop a simulation methodology for unbiased evaluation of \gls{ppp}-based large wireless networks, quantify the bias caused by insufficient simulation area and too few channel realizations, and use this methodology to compare the seven system types in detail.
      \item To interpret the numerical results, we introduce the metric of relative capacity and show how the value of cooperation depends on the \gls{ap} density, \gls{cpu} density, \gls{ue} density, and on whether sites are colocated or distributed.
\end{itemize}
Our numerical results show that stronger cooperation consistently improves performance, but the size of the gain depends strongly on the deployment regime.
In particular, distributed sites become increasingly attractive at higher \gls{ap} densities, low \gls{cpu} densities are important to preserve their advantage, and already user coordination can provide substantial gains before fully cell-free operation is needed.

\subsection{Paper Outline}
The remainder of this paper is organized as follows.
\autoref{sec:properties} introduces the generalized network model, formalizes scalability, and classifies the seven system types considered in this work.
\autoref{sec:system_model} presents the transmission model and derives unified uplink and downlink \gls{se} expressions.
\autoref{sec:processing} then describes the scalable signal processing and association procedures used for fair performance comparison across the different architectures.
\autoref{sec:accuracy_of_evaluation} discusses the simulation methodology, including how to obtain unbiased estimates and how to choose sufficiently accurate simulation parameters.
The numerical comparison of the seven system types is provided in \autoref{sec:results}, and \autoref{sec:conclusion} concludes the paper.

\textit{\textbf{Reproducible research:}} The simulation code and processed data used to generate the numerical results in this paper will be made publicly available by the authors, and a persistent repository link will be added in the final version.

\textit{\textbf{Notation:}} Boldface lowercase letters $\myvec{x}$ and boldface uppercase letters $\mymat{X}$ denote vectors and matrices, respectively.
Calligraphic letters $\myset{X}$ denote sets, $\mygraph{G}$ denotes the network graph and $\mathscr{P}$ denotes power sets.
The superscripts $\transp$ and $\herm$ denote transpose and Hermitian transpose, respectively.
Moreover, $\ex{\cdot}$ denotes expectation, while $\diag(\cdot)$, $\abs{\cdot}$, and $\norm{\cdot}$ denote the diagonal operator, absolute value, and Euclidean norm.

%% file: properties.tex
\section{General Network Properties} \label{sec:properties}
We define a generalized system model for fair comparison of cellular and cooperative massive \gls{mimo} systems.
We begin by defining the fundamental components of our model.

\begin{definition}[Network]\label{def:network}
    A wireless network is a graph $\mygraph{G} = (\myset{J} \cup \myset{L}, \myset{E} \cup \myset{F})$.
    The node set is the union of a set of \glspl{cpu}~$\myset{J}$ and a set of \glspl{ap}~$\myset{L}$.
    The edge set consists of midhaul links $\myset{E} \subseteq \myset{J} \times \myset{J}$ and fronthaul links $\myset{F} \subseteq \myset{L} \times \myset{J}$.
    For the fronthaul links, it holds that $\forall l \in \myset{L} \, \exists! j \in \myset{J}: (l, j) \in \myset{F}$, i.e., each \gls{ap} is connected to exactly one \gls{cpu}.
\end{definition}

\gls{ap} nodes represent antenna arrays connected to \glspl{cpu} by high-capacity, low-latency fronthaul links that allow extensive signaling.
\gls{cpu} nodes represent computational entities responsible for network tasks such as signal processing, resource allocation, and scheduling.
Midhaul links enable coordination between \glspl{cpu}, but may be more limited in capacity and latency than fronthaul links.
Each \gls{cpu} connects to the core network via a backhaul link, which is not explicitly modeled here.
To clarify the distinction between fronthaul and midhaul, we define:

\begin{definition}[Distributed/Colocated Site]
    The set of all \glspl{ap} $\myset{L}_j := \{l: (l,j) \in \myset{F}\}$ connected to a \gls{cpu}~$j \in \myset{J}$ via fronthaul links is called a site.
    If $\vert \myset{L}_j \vert > 1$ the site is called distributed, otherwise it is called colocated.
\end{definition}

Fronthaul links let the \gls{cpu} coordinate transmission and reception across all connected \glspl{ap}, effectively treating them as one antenna array, which we call a site.
Cooperation can also occur between sites:

\begin{definition}[Cooperation Group]
    The set $\myset{J}_j := \{j\} \cup \{j': (j,j') \in \myset{E}\}$ of a \gls{cpu} and its neighbors is called a cooperation group.
\end{definition}

The degree of inter-site cooperation depends on the midhaul capabilities discussed later in \autoref{subsec:network_classification}.
Building upon \autoref{def:network}, we define a specific operational context.

\begin{definition}[Scenario] \label{def:scenario}
    A scenario is a triplet $\mathcal{S} = (\mygraph{G}, \mathcal{K}, \Psi)$, where $\mygraph{G}$ is a network graph, $\mathcal{K}$ is a set of \glspl{ue}, and $\Psi$ is an association policy.
    The association policy is a mapping $\Psi: \mathcal{K} \rightarrow \myset{J} \times \powerset{\myset{L}} \times \powerset{\myset{L}} \times \powerset{\myset{L}}$, $k \mapsto (j^\star_k, \myset{L}^\mathrm{m}_k, \myset{L}^\mathrm{e}_k, \myset{L}^{\mathrm{s}}_k)$, i.e. it assigns each \gls{ue}~$k$:
    \begin{itemize}
        \item A master \gls{cpu}~$j^\star_k \in \myset{J}$
        \item A set of measuring \glspl{ap}~$\myset{L}^\mathrm{m}_k \subseteq \myset{L}$
        \item A set of estimating \glspl{ap}~$\myset{L}^\mathrm{e}_k \subseteq \myset{L}$
        \item A set of serving \glspl{ap}~$\myset{L}^\mathrm{s}_k \subseteq \myset{L}$
    \end{itemize}
\end{definition}

A scenario represents a complete snapshot of the network's state and configuration.
It combines the fixed physical deployed network $\mygraph{G}$ with a set of active \glspl{ue}~$\mathcal{K}$ and a policy $\Psi$ that determines how the \glspl{ue} are served by the network.
Each \gls{ue}~$k$ is associated to a single master \gls{cpu}~$j^\star_k$ which acts as a gateway to the core network.
Each \gls{ap}~$l$ in the measuring set $\myset{L}^\mathrm{m}_k$ measures the statistical \gls{csi} of \gls{ue}~$k$ (i.e., the correlation matrix $\mymat{R}_{kl}$ and large scale fading coefficient $\beta_{kl}$).
Each \gls{ap}~$l$ in the estimating set $\myset{L}^\mathrm{e}_k$ acquires an estimate $\widehat{\myvec{h}}_{kl} \approx \myvec{h}_{kl}$ of the instantaneous \gls{csi} for \gls{ue}~$k$.
\gls{ue}~$k$ is served only by \glspl{ap} in $\myset{L}^\mathrm{s}_k$.
This separation allows us to model \glspl{ap} that can cancel interference from a \gls{ue}~$k$ without serving it with payload data.
Typically, we have $\myset{L}^\mathrm{s}_k \subseteq \myset{L}^\mathrm{e}_k \subseteq \myset{L}^\mathrm{m}_k$.

For notational convenience, it is useful to also define the \gls{ue}-specific set of measuring \glspl{cpu} $\myset{J}_k^\mathrm{m} := \{j \in \myset{J}: \myset{L}_j \cap \myset{L}^\mathrm{m}_k \neq \emptyset\}$, the set of estimating \glspl{cpu} $\myset{J}_k^\mathrm{e} := \{j \in \myset{J}: \myset{L}_j \cap \myset{L}^\mathrm{e}_k \neq \emptyset\}$, and the set of serving \glspl{cpu} $\myset{J}_k^\mathrm{s} := \{j \in \myset{J}: \myset{L}_j \cap \myset{L}^\mathrm{s}_k \neq \emptyset\}$.
The sets above are defined from the perspective of a \gls{ue}~$k$, but implicitly given are the corresponding sets from the perspective of an \gls{ap}~$l$ as $\myset{K}^\mathrm{m}_l := \{k \in \mathcal{K}: l \in \myset{L}^\mathrm{m}_k\}$, $\myset{K}^\mathrm{e}_l := \{k \in \mathcal{K}: l \in \myset{L}^\mathrm{e}_k\}$, and $\myset{K}^\mathrm{s}_l := \{k \in \mathcal{K}: l \in \myset{L}^\mathrm{s}_k\}$.
Similarly, we have from the perspective of a \gls{cpu}~$j$ the sets $\myset{K}^\mathrm{m}_j := \bigcup_{l \in \myset{L}_j} \myset{K}^\mathrm{m}_l$, $\myset{K}^\mathrm{e}_j := \bigcup_{l \in \myset{L}_j} \myset{K}^\mathrm{e}_l$, and $\myset{K}^\mathrm{s}_j := \bigcup_{l \in \myset{L}_j} \myset{K}^\mathrm{s}_l$.

For practical feasibility, the association policy must match the network structure and link assumptions (see \autoref{subsec:network_classification}).
For example, without midhaul links, all \glspl{ap} serving \gls{ue}~$k$ must be directly connected to its master \gls{cpu}~$j^\star_k$ by fronthaul.

We seek to statistically analyze the performance of \glspl{ue} in scenarios.
To avoid finite-network artifacts, the sets of \glspl{ue}, \glspl{ap}, and \glspl{cpu} are generated by spatial point processes on an infinite plane.
Yet practical networks (and simulations) can support only finite computation and signaling per node and edge.
This requirement is known as \emph{scalability} in the cell-free literature \cite{interdonatoScalabilityAspectsCellFree2019,bjornsonScalableCellFreeMassive2020}, which we define adjusted to our graph model as follows.
\begin{definition}[Scalability]
    A scenario is called scalable if with probability 1:
    \begin{itemize}
        \item Each site~$j$ contains finitely many \glspl{ap}, i.e., $\forall j: \vert \myset{L}_j \vert < \infty$,
        \item Each \gls{cpu}~$j$ has finitely many neighboring \glspl{cpu}, i.e., $\forall j: \vert \myset{J}_j \vert < \infty$,
        \item Each \gls{ap}~$l$ is associated with a finite number of \glspl{ue}, i.e., $\forall l: \vert \myset{K}^\mathrm{m}_l \vert < \infty$, $\vert \myset{K}^\mathrm{e}_l \vert < \infty$, and $\vert \myset{K}^\mathrm{s}_l \vert < \infty$
    \end{itemize}
\end{definition}

\subsection{Network Classification} \label{subsec:network_classification}
\begin{figure*}[t]
    \centering
    \begin{subfigure}{0.25\textwidth}
        \includegraphics[width=\textwidth]{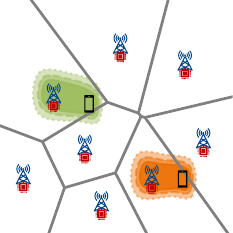}
        \caption{Conventional Cellular}
    \end{subfigure}%
    \begin{subfigure}{0.25\textwidth}
        \includegraphics[width=\textwidth]{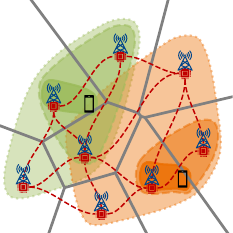}
        \caption{Coordinated Multi-Cell}
    \end{subfigure}%
    \begin{subfigure}{0.25\textwidth}
        \includegraphics[width=\textwidth]{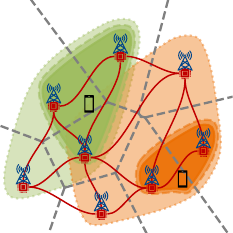}
        \caption{Distributed Cell-Free}
    \end{subfigure}%
    \begin{subfigure}{0.25\textwidth}
        \includegraphics[width=\textwidth]{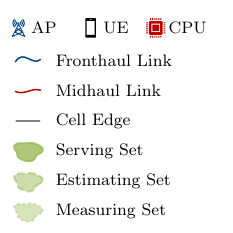}
        \caption*{\phantom{Legend}}
    \end{subfigure}
    \begin{subfigure}{0.25\textwidth}
        \includegraphics[width=\textwidth]{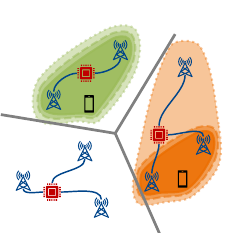}
        \caption{Cellular DAS}
    \end{subfigure}%
    \begin{subfigure}{0.25\textwidth}
        \includegraphics[width=\textwidth]{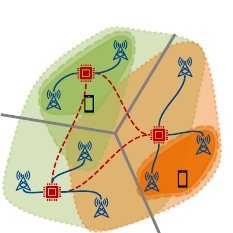}
        \caption{Coordinated DAS}
    \end{subfigure}%
    \begin{subfigure}{0.25\textwidth}
        \includegraphics[width=\textwidth]{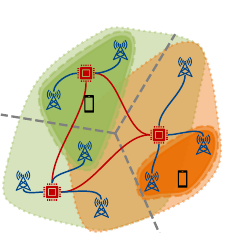}
        \caption{Hybrid Cell-Free}
    \end{subfigure}%
    \begin{subfigure}{0.25\textwidth}
        \includegraphics[width=\textwidth]{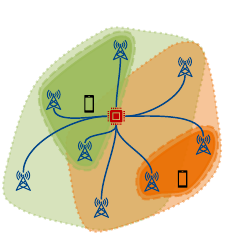}
        \caption{Centralized Cell-Free}
    \end{subfigure}%
    \caption{Considered system types. Upper row: Systems with \emph{colocated sites}. Lower row: Systems with \emph{distributed sites}. Midhaul capabilities from left to right are \emph{No Cooperation}, \emph{User Coordination}, \emph{Limited Joint Service}, \emph{Full Joint Service}.} \label{fig:system_types}
\end{figure*}

Equipped with the basic network properties, we now classify different types of networks that can be represented in our framework.
Consider four modes of inter-site cooperation through midhaul links:
\begin{enumerate}
	      \item {
	            \emph{\textbf{No Cooperation}}:
	          Midhaul links do not exist or are not used, hence no cooperation between sites is possible.
	          }
	      \item {
	            \emph{\textbf{User Coordination}}:
	          Midhaul links only exchange information about \gls{ue} presence.
	          A \gls{ue} is served by one site, but connected sites know that it exists.
	          They can therefore estimate local \gls{csi} to this \gls{ue} and manage interference with it when serving other \glspl{ue}.
	          This mode requires only small signaling capacity.
	          Midhaul latency must be moderate, on the order of $10 - \SI{100}{\milli\second}$, so that the \gls{ue} is still active when the information arrives.
	          }
	      \item {
	            \emph{\textbf{Limited Joint Service}}:
	          Midhaul links exchange payload data and statistical \gls{csi} about the \gls{ue}.
	          The \gls{ue} is jointly served by cooperating sites.
	          However, each site has access only to its local instantaneous \gls{csi} and must design its beamformer locally.
	          Beamforming vectors are therefore not coordinated across sites.
	          This mode requires large midhaul capacity, since sites exchange (soft) payload data.
	          Midhaul latency must be low to moderate, on the order of $\SI{10}{\milli\second}$, so that statistical \gls{csi} is not outdated on arrival.
	          }
	      \item {
	            \emph{\textbf{Full Joint Service}}:
	          Midhaul links exchange data and full \gls{csi}.
	          All cooperating sites then act as one large antenna array with one jointly designed beamformer.
	          This mode imposes the same midhaul requirements as fronthaul between \glspl{cpu} and \glspl{ap}.
	          It requires both high data rate and very low latency on the order of $\SI{1}{\milli\second}$, since instantaneous \gls{csi} changes quickly and must not be outdated on arrival.
	          }
\end{enumerate}

These cooperation modes can be mixed within one network, but we treat them as distinct categories for clarity.
Combining site type (colocated or distributed) with midhaul cooperation mode yields the seven network types in \autoref{fig:system_types}.

A network with colocated sites and no midhaul links corresponds to a \emph{conventional cellular} system where \glspl{ue} are served by non-cooperating base stations.
For each \gls{ue}~$k$, the association policy assigns a master \gls{cpu}~$j_k^\star$.
Without midhaul links, the cooperation group of \gls{ue}~$k$ is limited to its master \gls{cpu}, i.e., $\myset{J}_{j_k^\star} = \{j_k^\star\}$.
Since \gls{cpu}~$j_k^\star$ is connected to a single \gls{ap}~$l_{j_k^\star}$, the measuring, estimating and serving sets are $\myset{L}_{k}^\mathrm{m} = \myset{L}_{k}^\mathrm{e} = \myset{L}_{k}^\mathrm{s} = \{l_{j_k^\star}\}$.

We call a network with distributed sites and no midhaul links a \emph{cellular \gls{das}}.
Unlike in the conventional cellular case, the measuring, estimating and serving sets of \gls{ue}~$k$ may contain multiple \glspl{ap}, but each must be directly connected to the master \gls{cpu}~$j_k^\star$.
This architecture is closely related to a single \gls{cran} cluster \cite{wuCloudRadioAccess2015,schwarzDynamicDistributedAntenna2018,ghauchCoordinationAntennaDomain2016,chabbouhTwostageRRHClustering2017}.
There, so-called \glspl{rrh} with minimal processing capabilities are connected to a central \gls{bbu} that coordinates the transmission and reception.
In terms of signal processing, the \gls{bbu} treats the \glspl{rrh} as a single large antenna array with many antennas.
Typically, in works on \gls{cran}, scalability is not explored, but it can trivially be achieved by limiting the coverage area of a single \gls{bbu}, i.e., by deploying multiple \gls{cran} clusters (which are essentially cells) to cover a large area.

Midhaul links supporting \emph{user coordination} enable inter-cell interference management.
The cooperation group of \gls{ue}~$k$ may now include multiple \glspl{cpu}, so its measuring and estimating sets can extend to \glspl{ap} connected to any cooperating \gls{cpu}.
However, since no payload data is exchanged over midhaul, the serving set remains limited to \glspl{ap} connected to the master \gls{cpu}.
A colocated deployment in this configuration corresponds to a \emph{coordinated multi-cell} system, while a distributed deployment corresponds to a \emph{coordinated \gls{das}}.
Coordinated multi-cell systems have been studied extensively in the context of the coordinated beamforming mode of \gls{comp} \cite{leeCoordinatedMultipointTransmission2012,leeCoordinatedMultipointTransmission2012a,schwarzExploringCoordinatedMultipoint2014}.
Research on coordinated \gls{das} is more limited, but recent works in the cell-free context propose it as either a practical deployment strategy \cite{murakamiAnalysisCPUPlacement2022,ikamiCooperationMethodCPUs2023,tsukamotoFullStackTestbedInterSite2025} or a transitional step towards fully cell-free systems~\cite{femeniasCellsFreedom6Gs2025}.

The defining characteristic of a cell-free system is that each \gls{ue} can be served by a cluster of \glspl{ap} without being constrained by cell boundaries.
In our framework, this becomes possible once the midhaul supports payload data exchange.
With colocated sites and \emph{limited joint service}, the measuring and estimating sets are the same as in a coordinated multi-cell system, but the serving set may also include \glspl{ap} connected to other \glspl{cpu} in the cooperation group.
This architecture is commonly called \emph{distributed cell-free MIMO}, since most processing is local to the \glspl{cpu}.
It is the cell-free form requiring the least inter-site cooperation while still removing cell edges \cite{bjornsonMakingCellFreeMassive2020,leeFullyDistributedCellFree2024,schulzScalableCellFreeMassive2024}.

To exploit spatial diversity more fully, distributed sites can be combined with midhaul links that enable limited joint service.
This gives a \emph{hybrid cell-free MIMO system}, studied as a compromise between distributed scalability and joint-service gains \cite{schulzHybridReceiveCombining2025} and implementable using \gls{oran} technology \cite{ranjbarCellFreeMMIMOSupport2022,göttschFairnessSchedulingUserCentric2024,demirCellFreeMassiveMIMO2024,beertenMobileCellFreeMassive2025}.

Finally, with \emph{full joint service}, all processing for each \gls{ue} can be centralized at its master \gls{cpu}.
All \glspl{ap} connected to any \gls{cpu} in the cooperation group are then treated as one large antenna array with jointly optimized beamformers.
The distinction between colocated and distributed sites disappears, since the system operates as one large-scale array and all \glspl{cpu} act as one virtual cloud processor.
This architecture is known as \emph{centralized cell-free MIMO}, the canonical cell-free version, because most processing occurs at the central unit \cite{nayebiCellFreeMassiveMIMO2015,ngoCellFreeMassiveMIMO2017,bjornsonMakingCellFreeMassive2020}.
Equivalently, it can be interpreted as a \gls{cran} architecture with one physical \gls{bbu} coordinating all \glspl{rrh} in the network, but such a system is not scalable.

%% file: system_model.tex
\section{Transmission Model and Spectral Efficiency} \label{sec:system_model}
Having defined the different network types, we now introduce a transmission model that allows us to evaluate the performance of such networks in terms of achievable \gls{se} in a unified manner.
We restrict ourselves to the \emph{ergodic} performance of \emph{static} scenarios.
Hence, we assume a scenario as defined in \autoref{def:scenario} is given and aim to obtain an expression for the achievable ergodic \gls{se} per \gls{ue}.
Importantly, the expressions given in the following are valid for all considered system types.
For the same network graph, the system types perform differently due to the different sets of nodes $\myset{L}_k^\mathrm{m}$, $\myset{L}_k^\mathrm{e}$, and $\myset{L}_k^\mathrm{s}$ associated with each \gls{ue}~$k$ by the scenario's association policy.

For each pair of \gls{ap}~$l$ and \gls{ue}~$k$, there exists a wireless channel.
We use the block fading channel model, meaning that the channel remains constant over a coherence block of $\tau_c$ symbols, but changes independently between blocks.
The \glspl{ap} are each equipped with $N$ antennas, while the \glspl{ue} are single-antenna devices.
The channel from \gls{ap}~$l$ to \gls{ue}~$k$ is denoted by $\myvec{h}_{kl} \in \mathbb{C}^N$ and drawn from a correlated Rayleigh distribution as
\begin{equation} \label{eq:rayleigh_fading}
    \myvec{h}_{kl} \sim \mathcal{CN}(\myvec{0}, \beta_{kl} \mymat{R}_{kl}),
\end{equation}
where $\beta_{kl}$ is the large-scale fading coefficient and the spatial correlation matrix $\mymat{R}_{kl}$ is normalized so that $\text{tr}(\mymat{R}_{kl}) = N$.
We denote the large-scale effects of the channel, i.e. the matrix $\beta_{kl} \mymat{R}_{kl}$, by statistical \gls{csi}.
The small-scale variations of the channel, i.e. the realization of $\myvec{h}_{kl}$, are referred to as instantaneous \gls{csi}.

Since the true capacity of the \gls{mimo} interference channel is generally unknown, we make two non-optimal assumptions in the operation of the system to simplify the analysis:
\begin{enumerate}
    \item linear receive combining and transmit precoding instead of optimal non-linear processing,
    \item \gls{tdd} operation with pilot-based channel estimation only in the uplink and exploiting channel reciprocity in the downlink.
\end{enumerate}
Both assumptions are common practice in the massive \gls{mimo} literature.
The \gls{tdd} protocol divides the coherence block into three phases: a pilot phase of $\tau_p$ symbols, an uplink data transmission phase of $\tau_u$ symbols, and a downlink data transmission phase of $\tau_d$ symbols, such that $\tau_c = \tau_p + \tau_u + \tau_d$.

\subsection{Uplink Pilot Transmission Phase} \label{subsec:pilot_phase}
Since we consider the ergodic case, we can assume the respective local statistical \gls{csi} is measured at the \glspl{ap} and \glspl{ue} with perfect accuracy, but the instantaneous \gls{csi} needs to be estimated for each coherence block.
For this, during the pilot phase, each \gls{ue}~$k$ transmits one of $\tau_p$ orthogonal pilot sequences.
To obtain a channel estimate $\widehat{\myvec{h}}_{kl}$, the \gls{ap}~$l$ correlates its received signal with the pilot assigned to \gls{ue}~$k$ and applies the standard \gls{mmse} estimator, i.e., we have $\myvec{h}_{kl} = \widehat{\myvec{h}}_{kl} + \tilde{\myvec{h}}_{kl}$.
When using the unbiased \gls{mmse} estimator \cite[Sec.~4.2]{demirFoundationsUserCentricCellFree2022}, the estimation error $\tilde{\myvec{h}}_{kl}$ is Gaussian distributed with zero mean and covariance matrix
\begin{equation} \label{eq:est_error_cov}
    \begin{split}
        \mymat{C}_{kl} = \ex{\tilde{\myvec{h}}_{kl} \tilde{\myvec{h}}_{kl} \herm}
        = \beta_{kl}\mymat{R}_{kl} - \\ \eta_k \tau_p \beta_{kl}\mymat{R}_{kl} \left( \sum_{k' \in \myset{K}_{t_k}} \eta_{k'} \tau_p \beta_{k'l}\mymat{R}_{k'l} + \sigma^2_\mathrm{ul} \myvec{I} \right)^{-1} \beta_{kl}\mymat{R}_{kl}.
    \end{split}
\end{equation}
Here, $\eta_k$ is the transmit power of the pilot sent by \gls{ue}~$k$, $\sigma^2_\mathrm{ul}$ is the uplink noise variance, and $\myset{K}_{t_k}$ is the set of \glspl{ue} that transmit the same pilot as \gls{ue}~$k$.
In practice the channel estimation can either be performed locally at the \gls{ap} or at the \gls{cpu} connected to it.
In the former case, the \gls{ap} explicitly shares the channel estimates with the \gls{cpu} while in the latter case the \gls{ap} forwards the received pilot signals to the \gls{cpu} which then performs the channel estimation.
Assuming error-free fronthaul links with infinite capacity, both approaches are theoretically equivalent: since channels to different \glspl{ap} are modeled as uncorrelated, joint processing of pilot observations from multiple \glspl{ap} does not improve the estimate of $\myvec{h}_{kl}$ beyond the local observation at \gls{ap}~$l$.

\begin{remark} \label{rem:partial_estimates}
    The per-coherence-block estimation procedure described above is only performed for the limited set of \glspl{ue} $\myset{K}_l^{\mathrm{e}}$ for each \gls{ap}~$l$.
    For \glspl{ue} that are not in this set but in the measuring set $\myset{K}_l^{\mathrm{m}}$, the \gls{ap}~$l$ only measures long-term statistical \gls{csi} of these \glspl{ue} and sets their channel estimates to its expectation, i.e., $\widehat{\myvec{h}}_{kl} = \ex{\myvec{h}_{kl}}$ for $k \in \myset{K}_l^{\mathrm{m}}$.
    In the case of Rayleigh fading, we have $\ex{\myvec{h}_{kl}} = \myvec{0}$ and hence $\mymat{C}_{kl} = \beta_{kl}\mymat{R}_{kl}$.
    For all remaining \glspl{ue} $k \notin \myset{K}_l^{\mathrm{e}} \cup \myset{K}_l^{\mathrm{m}}$ the \gls{ap}~$l$ is unaware of any \gls{csi} and we formally set $\widehat{\myvec{h}}_{kl}=\myvec{0}$ and $\mymat{C}_{kl} = \mymat{0}$ to simplify notation in the following sections.
\end{remark}

\subsection{Uplink Data Transmission Phase} \label{subsec:uplink_phase}
During the uplink phase, each \gls{ue}~$k$ transmits a signal $s_k^\mathrm{ul}$ with the average power constraint $\ex{|s_k^\mathrm{ul}|^2} = p_k \leq P^\mathrm{max}_\mathrm{ul}$.
Hence, the received signal at \gls{ap}~$l$ is given by
\begin{equation}
    \myvec{y}_{l}^\mathrm{ul} = \sum_{k \in \myset{K}} \myvec{h}_{kl} s_k^\mathrm{ul} + \myvec{n}_l \in \mathbb{C}^N,
\end{equation}
where $\myvec{n}_l \sim \mathcal{CN}(\myvec{0}, \sigma^2_{\mathrm{ul}} \mymat{I})$ is the \gls{awgn} term at \gls{ap}~$l$.

The signal is then forwarded to the \gls{cpu}~$j$ connected to \gls{ap}~$l$ over the fronthaul link.
For \gls{ue}~$k$, the \gls{cpu}~$j$ collects signals from all \glspl{ap} that serve \gls{ue}~$k$ and are connected to it, for convenience we define this set as
\begin{equation}
    \myset{L}_{kj} = \myset{L}_k^{\mathrm{s}} \cap \myset{L}_j.
\end{equation}
The \gls{cpu}~$j$ can interpret all of these \glspl{ap} as a single large \gls{mimo} array.
Using the concatenated channel vector ${\myvec{h}_{kj} := [\myvec{h}_{kl}]_{l \in \myset{L}_{kj}}}$ and the concatenated noise vector ${\myvec{n}_{kj} := [\myvec{n}_l]_{l \in \myset{L}_{kj}}}$, we can write the relevant effective receive signal for \gls{ue}~$k$ at \gls{cpu}~$j$ as
\begin{equation}
    \myvec{y}_{kj}^\mathrm{ul} = [\myvec{y}_l^\mathrm{ul}]_{l \in \myset{L}_{kj}} = \sum_{k' \in \myset{K}} \myvec{h}_{k'j} s_{k'}^\mathrm{ul} +\myvec{n}_{kj} \in \mathbb{C}^{N \cdot \abs{\myset{L}_{kj}}}.
\end{equation}

The \gls{cpu}~$j$ then performs a local linear combining step for \gls{ue}~$k$, resulting in the local estimate
\begin{equation} \label{eq:ul_cpu_received}
    \begin{split}
        \widehat{s}_{kj}^\mathrm{ul} &= \myvec{v}_{kj}\herm \myvec{y}_{kj}^\mathrm{ul} = \sum_{k' \in \myset{K}} \myvec{v}_{kj}\herm \myvec{h}_{k'j} s_{k'}^\mathrm{ul} + \myvec{v}_{kj}\herm \myvec{n}_{kj} \\
        &= \sum_{k' \in \myset{K}} \myvec{v}_{kj}\herm \widehat{\myvec{h}}_{k'j} s_{k'}^\mathrm{ul} + \sum_{k' \in \myset{K}} \myvec{v}_{kj}\herm \tilde{\myvec{h}}_{k'j} s_{k'}^\mathrm{ul} + \myvec{v}_{kj}\herm \myvec{n}_{kj} \\
        &= \widehat{g}_{kkj} s_{k}^\mathrm{ul} + \sum_{\substack{k' \in \myset{K} \\ k'\neq k}} \widehat{g}_{kk'j} s_{k'}^\mathrm{ul} + \sum_{k' \in \myset{K}} \tilde{g}_{kk'j} s_{k'}^\mathrm{ul} + \tilde{n}_{kj}.
    \end{split}
\end{equation}
Hence, the concatenation of the wireless \gls{mimo} channel with the receive combining vector $g_{kkj} := \myvec{v}_{kj}\herm \myvec{h}_{kj}$  effectively becomes a \gls{siso} channel with estimated channel coefficient $\widehat{g}_{kkj} := \myvec{v}_{kj}\herm \widehat{\myvec{h}}_{kj}$, estimated interference through the estimated channels $\widehat{g}_{kk'j} := \myvec{v}_{kj}\herm \widehat{\myvec{h}}_{k'j}$ for $k' \in \myset{K}\setminus\{k\}$, residual interference due to channel estimation errors $\tilde{g}_{kk'j} := \myvec{v}_{kj}\herm \tilde{\myvec{h}}_{k'j}$ for $k' \in \myset{K}$ and a noise component $\tilde{n}_{kj} := \myvec{v}_{kj}\herm \myvec{n}_{kj} \sim \mathcal{CN}(0, \tilde{\sigma}^2_{\mathrm{ul}, j})$ with $\tilde{\sigma}^2_{\mathrm{ul}, j} = \sigma^2_{\mathrm{ul}} \norm{\myvec{v}_{kj}}^2$.

Note, that as described in \autoref{rem:partial_estimates}, the local channel estimates $\widehat{\myvec{h}}_{k'j}$ are only non-zero for finitely many \glspl{ue}~$k'$.
For \glspl{ue}~$k \notin \myset{K}_j^{\mathrm{e}} \cup \myset{K}_j^{\mathrm{m}}$ we have $\widehat{g}_{kk'j} = 0$ and therefore $\tilde{g}_{kk'j} = g_{kk'j}$.
Hence, the sum over $\widehat{g}_{kk'j}$ in \autoref{eq:ul_cpu_received} effectively is a finite sum and only a finite number of effective channel coefficients are used by the detector.
These effective channel coefficients are available at the \gls{cpu}~$j$ since it has access to both the local channel estimates $\widehat{\myvec{h}}_{k'j}$ and the local combining vector $\myvec{v}_{kj}$.

If \gls{ue}~$k$ is only served by a single \gls{cpu}, then the local estimate is also the final estimate and no further processing is needed.
This is always the case for systems where the midhaul links allow \emph{no cooperation} or only \emph{user coordination}.
Also, for the centralized cell-free \gls{mimo} system with \emph{full joint service}, each \gls{ue}~$k$ is served by the single cloud \gls{cpu} and again no further processing is needed.

However, the systems with \emph{limited joint service} (i.e., \emph{distributed} and \emph{hybrid cell-free} \gls{mimo}) require a two-stage detection scheme.
Here, we decompose the set of \glspl{cpu} serving \gls{ue}~$k$ into the master \gls{cpu}~$j^\star$ and $J_k$ cooperating \glspl{cpu}, i.e.,
\begin{equation}
    \myset{J}_k^{\mathrm{s}} = \{j^\star\} \cup \{j_1, \dots, j_{J_k}\}.
\end{equation}
The cooperating \glspl{cpu} send their local estimates for \gls{ue}~$k$ to the master \gls{cpu} where these are interpreted as a \gls{mimo} signal vector by forming the concatenation $[\widehat{s}^\mathrm{ul}_{kj}]_{j \in J_{k}^{\mathrm{s}}}.$
Finally, a linear fusion step is applied to obtain the final estimate
\begin{equation}
    \widehat{s}^\mathrm{ul}_k = \myvec{a}_k\herm [\widehat{s}^\mathrm{ul}_{kj}]_{j \in J_{k}^{\mathrm{s}}} = a_{k\star}^*\widehat{s}^\mathrm{ul}_{kj^\star} + \sum_{i=1}^{J_k} a_{i}^* \widehat{s}^\mathrm{ul}_{kj_i}.
\end{equation}
Note, that the case of a single serving \gls{cpu} is a special case of this expression by setting $J_k = 0$ and $a_{k\star}=1$.

Generally, there are two common ways to calculate the effective \gls{sinr} and bound the achievable ergodic uplink \gls{se} in \gls{mimo} systems with linear processing:
\begin{enumerate}
    \item Assuming that the detector has access to instantaneous \gls{csi}, the effective channel $g_{kkj}$ can be interpreted as a \emph{stochastic interference channel} leading to the classical bound given in \cite[Thm.~4.1]{björnsonMassiveMIMONetworks2017}
    \item Alternatively, if only statistical \gls{csi} is available at the detector, the known part of the effective channel becomes $\ex{g_{kkj}}$ and interpreting it as a \emph{deterministic interference channel} leads to the \gls{uatf} bound (also called hardening bound) given in \cite[Thm.~4.4]{björnsonMassiveMIMONetworks2017}.
\end{enumerate}
The former of the two bounds is tighter, but also has more stringent requirements on the \gls{csi} knowledge at the detector.
When the effective channel hardens sufficiently (i.e., when the effective channel behaves close to deterministic), the two bounds are close to each other \cite{björnsonMassiveMIMONetworks2017}.
However, in practice this is only the case for a large number of antennas, which is not necessarily true in our setting.
For example, the \gls{uatf} bound may severely underestimate the achievable \gls{se} of a small-cell system where each \gls{ue} is only served by a single \gls{ap} with few antennas \cite{bjornsonMakingCellFreeMassive2020}.
In the following we present a new bound that combines the two approaches to obtain a tight bound for our setting.

\begin{theorem}[Uplink \gls{se}] \label{thm:uplink_se}
    An achievable uplink ergodic \gls{se} of \gls{ue}~$k$ is given by
    \begin{equation} \label{eq:uplink_se}
        \mathrm{SE}_k^{\mathrm{ul}} = \frac{\tau_u}{\tau_c} \ex{\log_2(1 + \mathrm{SINR}_k^{\mathrm{ul}})},
    \end{equation}
    where $\mathrm{SINR}_k^\mathrm{ul} = $
    \begin{equation}
        \resizebox{1.0\linewidth}{!}{$
        \frac{p_k \abs{\myvec{a}_k\herm \widehat{\myvec{g}}_{kk}}^2}{\sum \limits_{\substack{k' \in \myset{K} \\ k' \neq k}} \!\! p_{k'} \abs{\myvec{a}_k\herm \widehat{\myvec{g}}_{kk'}}^2 \! + \myvec{a}_k \herm \left(\sum \limits_{k' \in \myset{K}} \!\! p_{k'} \ex{\tilde{\myvec{g}}_{kk'}\tilde{\myvec{g}}_{kk'}\herm} \!\! \right)\myvec{a}_k + \myvec{a}_k \herm \tilde{\mymat{\Sigma}}_\mathrm{ul}\myvec{a}_k}
        $}
    \end{equation}
    with
    \begin{equation}
        \widehat{\myvec{g}}_{kk'} =
        \begin{bmatrix}
            \widehat{g}_{kk'j^\star} \\
            \ex{g_{kk'j_1}}          \\
            \vdots                   \\
            \ex{g_{kk'j_{J_k}}}
        \end{bmatrix},
        \tilde{\myvec{g}}_{kk'} =
        \begin{bmatrix}
            \tilde{g}_{kk'j^\star}       \\
            g_{kk'j_1} - \ex{g_{kk'j_1}} \\
            \vdots                       \\
            g_{kk'j_{J_k}} - \ex{g_{kk'j_{J_k}}}
        \end{bmatrix}
    \end{equation}
    and $\tilde{\myvec{\Sigma}}_\mathrm{ul} = \diag\left([\tilde{\sigma}^2_{\mathrm{ul}, j}]_{j \in \myset{J}_k^{\mathrm{s}}}\right)$.
    The expectations are taken with respect to the channel realizations.
    \begin{proof}[Proof sketch]
        Interpret $\myvec{a}_k\herm \widehat{\myvec{g}}_{kk}$ as a scalar stochastic interference channel, apply Shannon's capacity formula and consider that only the fraction ${\tau_u}/{\tau_c}$ is used for uplink data transmission.
    \end{proof}
\end{theorem}
The key difference to the classical bounds is that the known part of the effective channel is chosen component-wise according to the \gls{csi} available at the detector: instantaneous effective-channel estimates are used for the master \gls{cpu}, while only statistical means are used for the cooperating \glspl{cpu}.
If \gls{ue}~$k$ is only served by a single \gls{cpu}, then we can set $a_{k\star} = 1$ and we have $\widehat{\myvec{g}}_{kk'} = \widehat{g}_{kk'j^\star}$ and $\tilde{\myvec{g}}_{kk'} = \tilde{g}_{kk'j^\star}$.
Then, the above expression simplifies to the classical bound given in \cite[Thm.~4.1]{björnsonMassiveMIMONetworks2017}.
On the other hand, if one assumes that the detector has only statistical \gls{csi} available for all \glspl{cpu}~$j \in \myset{J}_k^{\mathrm{s}}$, then one gets $\widehat{\myvec{g}}_{kk'} = [\ex{\widehat{g}_{kk'j}}]_{j \in \myset{J}_k^{\mathrm{s}}}$ and $\tilde{\myvec{g}}_{kk'} = [g_{kk'j} - \ex{g_{kk'j}}]_{j \in \myset{J}_k^{\mathrm{s}}}$ and the expression simplifies to the \gls{uatf} bound given in \cite[Thm.~4.4]{björnsonMassiveMIMONetworks2017} (note that in this case $\mathrm{SINR}_k^\mathrm{ul}$ is deterministic and the expectation from \autoref{eq:uplink_se} can be omitted).

\subsection{Downlink Data Transmission Phase} \label{subsec:downlink_phase}
In the first step of the downlink phase, the signal $s_k^\mathrm{dl}$ with unit power $\norm{s_k^\mathrm{dl}}^2 = 1$ is distributed from the master \gls{cpu}~$j^\star$ to all cooperating \glspl{cpu}~$j \in \myset{J}_k^{\mathrm{s}}$ over the midhaul links.
If \gls{ue}~$k$ is only served by a single \gls{cpu}, this step is skipped.

Analogously to the uplink, \gls{cpu}~$j$ interprets the set $\myset{L}_{kj}$ of all \glspl{ap} connected to it that serve \gls{ue}~$k$ as a single large \gls{mimo} array with channel $\myvec{h}_{kj} := [\myvec{h}_{kl}]_{l \in \myset{L}_{kj}}$.
The \gls{cpu}~$j$ then designs a linear precoding vector $\myvec{w}_{kj} = [\myvec{w}_{kl}]_{l \in \myset{L}_{kj}} \in \mathbb{C}^{N \cdot \abs{\myset{L}_{kj}}}$ with average unit norm $\ex{\norm{\myvec{w}_{kj}}^2} = 1$ for \gls{ue}~$k$.
Further, the \gls{cpu}~$j$ allocates a transmit power $\rho_{kj}$ to \gls{ue}~$k$.
Hence, the part of the signal sent from \gls{cpu}~$j$ intended for \gls{ue}~$k$ is given by
\begin{equation}
    \myvec{x}_{kj} = \sqrt{\rho_{kj}} \myvec{w}_{kj} s_k^\mathrm{dl} \in \mathbb{C}^{N \cdot \abs{\myset{L}_{kj}}}.
\end{equation}

The received signal at \gls{ue}~$k$ is given by the superposition of the transmit signals from all \glspl{cpu}, i.e.,
\begin{equation}
    \begin{split}
        y_k^\mathrm{dl} &= \sum_{k' \in \myset{K}} \sum_{j \in \myset{J}_{k'}^\mathrm{s}} \sqrt{\rho_{k'j}} \myvec{h}_{kj}\herm \myvec{w}_{k'j} s_{k'}^\mathrm{dl} + n_{k} \\
        &= \sum_{k' \in \myset{K}} \sum_{j \in \myset{J}_{k'}^\mathrm{s}} \sqrt{\rho_{k'j}} f_{kk'j} s_{k'}^\mathrm{dl} + n_{k},
    \end{split}
\end{equation}
meaning effectively we get a \gls{siso} channel with coefficient $f_{kk'j} := \myvec{h}_{kj}\herm \myvec{w}_{k'j}$.

Importantly, the transmit signal is subject to per-\gls{ap} power constraints.
Concretely, consider the transmit signal from the perspective of an \gls{ap}~$l$:
For each \gls{ue}~$k'$ served by \gls{ap}~$l$, the \gls{ap}~$l$ transmits the subvector $\myvec{w}_{k'l}$ of the precoding vector $\myvec{w}_{k'j}$ designed at the \gls{cpu}~$j$ connected to \gls{ap}~$l$.
Hence, the part of the downlink transmit signal sent from \gls{ap}~$l$ is given by
\begin{equation}
    \myvec{x}_{l} = \sum_{k' \in \myset{K}_l^{\mathrm{s}}} \sqrt{\rho_{k'j}} \myvec{w}_{k'l} s_{k'}^\mathrm{dl} \in \mathbb{C}^N.
\end{equation}
Therefore, the precoding vectors and power allocation coefficients need to be designed at \gls{cpu}~$j$ such that the average power constraint
\begin{equation} \label{eq:dl_power_constraint}
    \ex{\norm{\myvec{x}_l}^2} = \sum_{k' \in \myset{K}_l^{\mathrm{s}}} \rho_{k'j} \ex{\norm{\myvec{w}_{k'l}}^2} \leq P^\mathrm{max}_\mathrm{dl}
\end{equation}
is satisfied at each \gls{ap}~$l \in \myset{L}_j$.

Contrary to the uplink, the \gls{ue}~$k$ only has access to statistical \gls{csi} in the downlink, and we bound the downlink \gls{se} using the \gls{uatf} bound \cite[Thm. 6.1]{demirFoundationsUserCentricCellFree2022}.

\begin{theorem}[Downlink \gls{se}] \label{thm:downlink_se}
    An achievable downlink ergodic \gls{se} of \gls{ue}~$k$ is given by
    \begin{equation} \label{eq:downlink_se}
        \mathrm{SE}_k^{\mathrm{dl}} = \frac{\tau_d}{\tau_c} \log_2(1 + \mathrm{SINR}_k^{\mathrm{dl}}),
    \end{equation}
    where $\mathrm{SINR}_k^\mathrm{dl} = $
    \begin{equation}
        \frac{\abs{\myvec{\mu}_k \transp \ex{\myvec{f}_{kk}}}^2}{\sum_{k' \in \myset{K}} \myvec{\mu}_{k'}\transp\ex{\myvec{f}_{kk'}\myvec{f}_{kk'}\herm}\myvec{\mu}_{k'} - \abs{\myvec{\mu}_k \transp \ex{\myvec{f}_{kk}}}^2 + \sigma^2_{\mathrm{dl}}}
    \end{equation}
    with $\myvec{f}_{kk'} = [f_{kk'j}]_{j \in \myset{J}_{k'}^{\mathrm{s}}}$ and $\myvec{\mu}_k = [\sqrt{\rho_{kj}}]_{j \in \myset{J}_{k}^{\mathrm{s}}}$.
    The expectations are taken with respect to the channel realizations.
    \begin{proof}[Proof sketch]
        Interpret $\myvec{\mu}_k \transp \ex{\myvec{f}_{kk}}$ as a scalar deterministic interference channel, apply Shannon's capacity formula and consider that only the fraction ${\tau_d}/{\tau_c}$ is used for downlink data transmission.
    \end{proof}
\end{theorem}

%% file: signal_processing.tex
\section{Spectrally Efficient Scalable Processing} \label{sec:processing}
In this section we describe the signal processing schemes that are used to evaluate the performance of the considered systems.
Our objective is to assess the maximally achievable \gls{se} under the architectural constraints of each system. Since we focus on scalable deployments, only local \gls{csi} is assumed to be available at each \gls{cpu}. As a result, globally optimal processing strategies that rely on network-wide instantaneous \gls{csi} are not feasible.

To enable fair comparison, we adopt processing schemes from the literature that are known to perform well, even if they are computationally demanding in practice.
This avoids underestimating the achievable performance of each system.
For example, we employ a generous user-centric clustering algorithm where each \gls{ap} is allowed to serve many \glspl{ue}.
Such clustering improves performance compared to restricting service only to the very strongest \gls{ap}-\gls{ue} links, at the cost of increased complexity.
Similarly, in the uplink we use local \gls{mmse} combining, even though an alternative scheme such as \gls{rzf} may achieve a better complexity-performance trade-off by reducing the computational complexity at the cost of a small performance loss \cite{jiangWhatMostEfficient2025}.

A systematic investigation of trade-offs between complexity and performance for the different systems is beyond the scope of this work and is left for future research.
The focus here is on establishing realistic upper performance bounds under scalable and locally informed processing.
The following subsections provide details on initial access, pilot assignment, and the association policy, followed by the specific signal processing approaches employed in the uplink and downlink.

\subsection{Initial Access, Pilot Assignment and Association Policy} \label{subsec:initial_access}
As outlined in \autoref{subsec:network_classification}, the network type determines the admissible \gls{ap} associations of each \gls{ue}~$k$ through the sets $\myset{L}^{\mathrm{m}}_k, \myset{L}^{\mathrm{e}}_k, \myset{L}^{\mathrm{s}}_k$.
We use a joint procedure for initial access, pilot assignment, and \gls{ap}-\gls{ue} association that adapts to all seven systems.
Determining the set of serving \glspl{ap} $\myset{L}^\mathrm{s}_k$ is commonly called \gls{dcc} and different approaches exist \cite{interdonatoScalabilityAspectsCellFree2019,chenStructuredMassiveAccess2021,wangClusteredCellFreeNetworking2023,mussbahPilotContaminationReduction2023}.
Our scheme adapts \cite{bjornsonScalableCellFreeMassive2020} to the multi-\gls{cpu} model (as in \cite{schulzHybridReceiveCombining2025}) and introduces \emph{measuring} sets to separate \gls{csi} acquisition levels and keep statistical-\gls{csi} handling scalable.

\begin{algorithm}[t]
    \caption{Joint initial access, pilot assignment, and association for a new \gls{ue}~$k$} \label{alg:association_policy}
    \begin{algorithmic}[1]
        \Require $t_{k'}$ for all \glspl{ue}~$k'$ already connected \\
        current $\myset{K}^{\mathrm{m}}_l, \myset{K}^{\mathrm{e}}_l, \myset{K}^{\mathrm{s}}_l$ for all \glspl{ap}~$l$
        \State $l^\star \gets \argmax_{l \in \myset{L}} \left\{ \beta_{kl} \right\}$
        \State $j^\star \gets$ \gls{cpu} connected to \gls{ap}~$l^\star$
        \State $t_k \gets \argmin_{t \in \{1, \dots, \tau_p\}} \sum_{k' \in \myset{K}^{\mathrm{m}}_{l^\star}:t_{k'} = t} \beta_{k'l^\star}$
        \For{each \gls{ap}~$l \in \myset{L}_{j^\star}$}
        \If{$\beta_{kl}>\beta^\mathrm{m}$}
        \State $\myset{K}^{\mathrm{m}}_l \leftarrow \myset{K}^{\mathrm{m}}_l \cup \{k\}$
        \EndIf
        \If{$\beta_{kl}>\beta^\mathrm{e}$}
        \State $\myset{K}^{\mathrm{e}}_l \leftarrow \myset{K}^{\mathrm{e}}_l \cup \{k\}$
        \EndIf
        \If{$\beta_{kl}>\beta^\mathrm{c}$}
        \If{$\exists k' \in \myset{K}^{\mathrm{s}}_l: t_k = t_{k'}$}
        \If{$\beta_{kl} > \beta_{k'l}$}
        \State $\myset{K}^{\mathrm{s}}_l \leftarrow \myset{K}^{\mathrm{s}}_l \backslash \{k'\}$
        \State $\myset{K}^{\mathrm{s}}_l \leftarrow \myset{K}^{\mathrm{s}}_l \cup \{k\}$
        \EndIf
        \Else
        \State $\myset{K}^{\mathrm{s}}_l \leftarrow \myset{K}^{\mathrm{s}}_l \cup \{k\}$
        \EndIf
        \EndIf
        \EndFor
        \If{\emph{user coordination} allowed}
        \For{each \gls{cpu}~$j$ connected to $j^\star$}
        \For{each \gls{ap}~$l \in \myset{L}_{j}$}
        \If{$\beta_{kl}>\beta^\mathrm{m}$}
        \State $\myset{K}^{\mathrm{m}}_l \leftarrow \myset{K}^{\mathrm{m}}_l \cup \{k\}$
        \EndIf
        \If{$\beta_{kl}>\beta^\mathrm{e}$}
        \State $\myset{K}^{\mathrm{e}}_l \leftarrow \myset{K}^{\mathrm{e}}_l \cup \{k\}$
        \EndIf
        \EndFor
        \EndFor
        \EndIf
        \If{\emph{joint service} allowed}
        \For{each \gls{cpu}~$j$ connected to $j^\star$}
        \For{each \gls{ap}~$l \in \myset{L}_{j}$}
        \If{$\beta_{kl}>\beta^\mathrm{c}$}
        \If{$\exists k' \in \myset{K}^{\mathrm{s}}_l: t_k = t_{k'}$}
        \If{$\beta_{kl} > \beta_{k'l}$}
        \State $\myset{K}^{\mathrm{s}}_l \leftarrow \myset{K}^{\mathrm{s}}_l \backslash \{k'\}$
        \State $\myset{K}^{\mathrm{s}}_l \leftarrow \myset{K}^{\mathrm{s}}_l \cup \{k\}$
        \EndIf
        \Else
        \State $\myset{K}^{\mathrm{s}}_l \leftarrow \myset{K}^{\mathrm{s}}_l \cup \{k\}$
        \EndIf
        \EndIf
        \EndFor
        \EndFor
        \EndIf
    \end{algorithmic}
\end{algorithm}
\autoref{alg:association_policy} is applied whenever a new \gls{ue}~$k$ connects.
For initial access, \gls{ue}~$k$ searches for periodically transmitted synchronization signals, selects the \gls{ap}~$l^\star$ with the strongest channel as its \emph{master} \gls{ap}, and performs a standard random access procedure to establish contact and synchronization with it.
The connected \gls{cpu}~$j^\star$ becomes its master \gls{cpu}.
The pilot index $t_k$ is chosen to minimize known local pilot contamination at $l^\star$ by comparing $\beta_{kl^\star}$ with the large-scale fading coefficients of already connected \glspl{ue} in $\myset{K}^{\mathrm{m}}_{l^\star}$.

The \glspl{ap} connected to $j^\star$ add \gls{ue}~$k$ to $\myset{K}^{\mathrm{m}}_l$ and $\myset{K}^{\mathrm{e}}_l$ whenever $\beta_{kl}$ exceeds $\beta_\mathrm{m}$ and $\beta_\mathrm{e}$, respectively.
They add \gls{ue}~$k$ to $\myset{K}^{\mathrm{s}}_l$ if $\beta_{kl}>\beta_\mathrm{c}$ and no stronger \gls{ue}~$k'$ with the same pilot is already served by \gls{ap}~$l$.
If \emph{user coordination} is allowed, neighboring \glspl{cpu} repeat the measuring and estimation updates for their \glspl{ap}.
If \emph{joint service} is allowed, they also apply the same serving-set update.

Repeating this procedure for each new \gls{ue} keeps $\myset{L}^\mathrm{m}_k$ and $\myset{K}^{\mathrm{m}}_l$ finite and thus the system scalable.
All steps use only local information at each \gls{cpu} and \gls{ap} and can be implemented in distributed fashion.
In practice, initial access measures synchronization-signal \gls{rsrp} rather than $\beta_{kl}$ directly, so $\beta_\mathrm{m}, \beta_\mathrm{e}, \beta_\mathrm{c}$ can be translated to corresponding \gls{rsrp} thresholds.
With effective signaling and power control, smaller thresholds can improve performance by increasing the sizes of $\myset{L}^\mathrm{m}_k, \myset{L}^\mathrm{e}_k, \myset{L}^\mathrm{s}_k$ at the cost of higher complexity, until gains saturate for very weak channels.
To avoid underestimating achievable performance, we use very low thresholds: $\beta_\mathrm{m}=\qty{-140}{\deci\bel}$, $\beta_\mathrm{e}=\qty{-120}{\deci\bel}$, and $\beta_\mathrm{c}=\qty{-120}{\deci\bel}$, corresponding with a transmit power of \qty{20}{\dBm} to \glspl{rsrp} of \qty{-120}{\dBm} (very weak, but measurable) and \qty{-100}{\dBm} (weak, but practical), respectively.

\subsection{Uplink Combining and Power Control} \label{subsec:uplink_processing}
We presented the general uplink signal model in \autoref{subsec:uplink_phase} resulting in the \gls{se} expression given in \autoref{thm:uplink_se}.
Now, we describe how to find combining vectors $\myvec{v}_{kj}$, fusion weights $a_{kj}$ and transmit powers $p_k$ that maximize this \gls{se}.

The combining vector $\myvec{v}_{kj}$ is locally designed at \gls{cpu}~$j$ and we choose the \gls{lmmse} combining scheme proposed in \cite{bjornsonMakingCellFreeMassive2020} which minimizes the conditional mean square error given the channel estimates $\widehat{\myvec{h}}_{k'j}$, i.e., using our notation from \autoref{eq:ul_cpu_received}
\begin{equation} \label{eq:lmmse_ansatz}
    \myvec{v}_{kj} = \argmin_{\myvec{v}} \ex{\abs{\myvec{v}\herm \myvec{y}_{kj}^\mathrm{ul} - s_k^\mathrm{ul}}^2 \mid \{\widehat{\myvec{h}}_{k'j}\}_{k' \in \myset{K}}}.
\end{equation}
To ensure scalability, we employ the partial version of the scheme \cite{nayebiPerformanceCellfreeMassive2016,bjornsonScalableCellFreeMassive2020}, where instead of all \glspl{ue}~$k' \in \myset{K}$ only a finite number of interfering \glspl{ue} is considered in the computation of the combiner.
Concretely, we set the considered set of interferers to the set of \glspl{ue} $\myset{K}^{\mathrm{m}}_j$ of which the \gls{cpu}~$j$ has access to statistical \gls{csi} and we get
\begin{equation} \label{eq:uplink_combiner}
    \myvec{v}_{kj} = p_k \left( \sum_{k' \in \myset{K}^{\mathrm{m}}_j} p_{k'}(\widehat{\myvec{h}}_{k'j} \widehat{\myvec{h}}_{k'j}\herm + \mymat{C}_{k'j}) + \sigma^2_\mathrm{ul} \mymat{I} \right)^{-1} \widehat{\myvec{h}}_{kj},
\end{equation}
where $\mymat{C}_{k'j}$ is the block-diagonal matrix obtained by concatenating the individual estimation error covariance matrices $\mymat{C}_{k'l}$ from \autoref{eq:est_error_cov} for $l \in \myset{L}_{k'j}$.
Note, that this combining scheme requires only local \gls{csi}, but no \gls{csi} of \glspl{ap} connected to other \glspl{cpu}.
For those subvectors $\myvec{h}_{k'l}$ of the channel estimates $\myvec{h}_{k'j}$ where $l \notin \myset{L}^{\mathrm{e}}_{k'}$, i.e., the instantaneous channel is not estimated, the corresponding subvector of the channel estimate is given as described in \autoref{rem:partial_estimates}, so \autoref{eq:uplink_combiner} is well-defined.

For the second stage of the uplink detector, we again adopt the \gls{mmse} criterion in order to maximize the \gls{se} expression in \autoref{thm:uplink_se}.
Hence, we get
\begin{equation} \label{eq:uplink_fusion}
    \resizebox{1.0\linewidth}{!}{$
    \myvec{a}_k \! = \! p_k \! \left( \sum_{k' \in \myset{K}^{\mathrm{m}}_{j^\star}} p_{k'}(\widehat{\myvec{g}}_{kk'} \widehat{\myvec{g}}_{kk'}\herm \! + \! \ex{\tilde{\myvec{g}}_{kk'}\tilde{\myvec{g}}_{kk'}\herm}) \! + \! \tilde{\mymat{\Sigma}}_\mathrm{ul} \right)^{-1} \!\!\! \widehat{\myvec{g}}_{kk}.
    $}
\end{equation}
\begin{remark}
    In the context of distributed cell-free \gls{mimo}, usually a two-stage uplink detector with local combining at each \gls{ap} followed by \gls{lsfd} at the \gls{cpu} is used \cite[Ch.~5]{demirFoundationsUserCentricCellFree2022}.
    The optimization target of \gls{lsfd} is to maximize the \gls{uatf} bound and only statistical \gls{csi} is required.
    Our approach differs by also incorporating instantaneous \gls{csi} and by maximizing our newly derived \gls{se} expression in \autoref{thm:uplink_se}.
    This leads to closer to optimal performance, but requires the fusion weights to be updated for every coherence block instead of only when the large-scale fading coefficients change.
\end{remark}
For uplink power control, we consider full power transmission, i.e., $p_k = P^\mathrm{max}_\mathrm{ul}, \forall k \in \myset{K}$, since this scheme is known to roughly approximate the maximum achievable sum \gls{se} in the network \cite[Ch.~7]{demirFoundationsUserCentricCellFree2022}.

\subsection{Downlink Precoding and Power Allocation} \label{subsec:downlink_processing}
As described in \autoref{subsec:downlink_phase}, in the downlink the precoding vectors $\myvec{w}_{kj}$ need to be designed together with the power allocation coefficients $\rho_{kj}$ in a way that does not violate the per-\gls{ap} power constraints in \autoref{eq:dl_power_constraint}.
Motivated by the uplink-downlink duality we simply reuse the uplink combining directions as precoding vectors by normalizing the combining vectors from \autoref{eq:uplink_combiner} to unit power, i.e.,
\begin{equation} \label{eq:downlink_precoder}
    \myvec{w}_{kj} = \frac{\myvec{v}_{kj}}{\sqrt{\ex{\norm{\myvec{v}_{kj}}^2}}}.
\end{equation}

For the power allocation, we use a local heuristic that lets each \gls{cpu} use its available per-\gls{ap} power budget, while assigning more power to \gls{ue}-\gls{cpu} pairs for which the \gls{cpu} is expected to contribute more strongly.
We measure this long-term contribution from the expected uplink fusion weights as
\begin{equation}
    \omega_{kj} =
    \frac{\abs{\ex{a_{kj}}}^2}{\sum_{j' \in \myset{J}^\mathrm{s}_k} \abs{\ex{a_{kj'}}}^2},
    \quad j \in \myset{J}^\mathrm{s}_k,
\end{equation}
and $\omega_{kj}=0$ otherwise.
The intuition is that a large fusion weight indicates that the local estimate from \gls{cpu}~$j$ is useful in the uplink detector for \gls{ue}~$k$ and by reciprocity, the same link is also a useful downlink transmission point.
Each \gls{cpu} can then compute using only local information the coefficients
\begin{equation}
    \bar{\rho}_{kj}
    = \omega_{kj}
    \frac{P^\mathrm{max}_\mathrm{dl}}{
        \max_{l \in \myset{L}_j}
        \sum_{k' \in \myset{K}^\mathrm{s}_l} \omega_{k'j}\ex{\norm{\myvec{w}_{k'l}}^2}},
    \quad j \in \myset{J}^\mathrm{s}_k.
\end{equation}
Thus, each \gls{cpu} performs a weighted equal-power allocation over its served \glspl{ue}, scaled so that the most heavily loaded connected \gls{ap} satisfies its power constraint.
The weights reduce power spent on \gls{cpu}-\gls{ue} links that contribute little to the joint transmission.

For systems where multiple \glspl{cpu} can serve the same \gls{ue}, it is possible that some \glspl{cpu} only serve a few \glspl{ue} and therefore can transmit with a very high power without violating the power constraints.
This can increase interference without providing a proportional useful-signal gain when the corresponding \gls{cpu}-\gls{ue} link has low importance.

Hence, in a second step, all \glspl{cpu} exchange their computed preliminary power coefficients $\bar{\rho}_{kj}$ and weights $\omega_{kj}$ with their neighbors and limit the power transmitted relative to the most important \gls{cpu} $j_k^\dagger=\argmax_{j \in \myset{J}^\mathrm{s}_k}\omega_{kj}$ by setting
\begin{equation}
    \rho_{kj}
    = \bar{\rho}_{kj}
    \min\left\{
    1,
    \frac{\omega_{kj}}{\omega_{k j_k^\dagger}}
    \frac{\bar{\rho}_{k j_k^\dagger}}{\bar{\rho}_{kj}}
    \right\}.
\end{equation}
We found both the importance weighting and interference control necessary to achieve good performance in the downlink, especially for weak \glspl{ue}.
The signaling necessary for interference control is negligible compared to the exchange of data and statistical \gls{csi} that is anyway done in systems where multiple \glspl{cpu} can serve the same \gls{ue}. 

%% file: eval_accuracy.tex
\section{Simulation Methodology} \label{sec:accuracy_of_evaluation}
Our goal is to estimate the unconditional \gls{cdf} of the uplink \gls{se} of an arbitrary \gls{ue} in an infinitely large network for each system type in \autoref{sec:properties}.
Since simulations are finite, we use Monte Carlo simulation and choose the simulation parameters such that finite-size bias is negligible.
To the best of our knowledge, their impact on accuracy has not been systematically discussed in the literature.
We first present an unbiased \gls{ppp}-based estimator using weighted \gls{se} samples in \autoref{subsec:monte_carlo_method}, then quantify the required simulation size in \autoref{subsec:needed_sim_size}.

\subsection{Unbiased Monte Carlo Simulation of Poisson Point Process Based Networks} \label{subsec:monte_carlo_method}
We place \glspl{ue} in a square area of size~$A$ according to a \gls{ppp} with density~$\lambda_K$, by drawing $K \sim \mathrm{Poisson}(\lambda_K A)$ and then sampling $K$ independent uniform positions.
\glspl{ap} and, for distributed-site systems, \glspl{cpu} are generated analogously with densities~$\lambda_L$ and~$\lambda_J$.
Fronthaul links connect each \gls{ap} to its nearest \gls{cpu}, while midhaul links follow the Delaunay triangulation of the \gls{cpu} positions.
The association policy in \autoref{alg:association_policy} assigns each \gls{ue}~$k$ a master \gls{cpu} and the sets $\myset{L}_k^\mathrm{m}$, $\myset{L}_k^\mathrm{e}$, and $\myset{L}_k^\mathrm{s}$, completing the scenario $\mathcal{S}$.
For the Rayleigh fading model in \autoref{eq:rayleigh_fading}, we draw $O$ independent realizations of each channel vector $\myvec{h}_{kl}$ and estimate all expectations in the \gls{se} expressions by sample means.

Toroidal (wrap-around) boundary conditions avoid boundary effects, so every \gls{ue} experiences a statistically equivalent environment.
Thus, we evaluate all \glspl{ue} in each scenario instead of only a single center \gls{ue}, which improves sampling efficiency \cite{fastenbauerInvestigationWraparoundTechniques2019}.
However, naively averaging over all \gls{se} samples for $Q$ scenarios gives a biased estimator of the \gls{cdf} $F(\mathrm{SE})$.
This is because for a scenario $q$, the number of \glspl{ue} $K_q$ is a realization of a Poisson distributed random variable, and the $K_q$ \gls{se} samples from that scenario follow the conditional distribution given $K_q$, not the desired unconditional distribution of a \emph{typical} \gls{ue}.
Hence, we weight the samples and get
\begin{equation} \label{eq:weighted_cdf}
    \widehat{F}(\mathrm{SE}) = \frac{\sum_{q=1}^{Q}\sum_{k=1}^{K_q} \mathbbm{1}(\mathrm{SE}_{q,k} \leq x)}{\sum_{q=1}^{Q} K_q},
\end{equation}
where $\mathbbm{1}$ represents the indicator function and $\mathrm{SE}_{q,k}$ is the \gls{se} of \gls{ue}~$k$ in scenario $q$.
This is especially important in the tails: scenarios with many \glspl{ue} usually yield lower per-\gls{ue} \gls{se} and more samples, and would otherwise be overrepresented.
This \emph{ratio estimator} is consistent, i.e., it converges to the true \gls{cdf} as $Q \to \infty$, and we use $Q=1000$ which gives a very low variance and remaining bias of the estimate.

\subsection{Needed Simulation Size for Accurate Evaluation} \label{subsec:needed_sim_size}
Too small simulation area sizes $A$ and too few channel realizations $O$ introduce bias in the estimated \gls{se} \gls{cdf}, but choosing them too large results in simulations with excessive computational complexity.
We systematically find suitable values by comparing the estimated \glspl{se} with computationally expensive reference simulations using very large $A$ and $O$, chosen such that further increases change the results only negligibly.
Note, that other simulation parameters, such as for example the pathloss and fading model, heavily influence the required values for $A$ and $O$.
We use standard parameters from the literature, as summarized in \autoref{tab:simulation_parameters}.
The introduced bias also differs for the type of simulated system.
We focus on the uplink of the three representative systems: cellular, distributed cell-free, and centralized cell-free.
Similar conclusions hold for the remaining system types and the downlink.

\begin{table}[t]
    \centering
    \caption{Default simulation parameters.}
    \label{tab:simulation_parameters}
    \begin{tabular*}{\columnwidth}{@{\extracolsep{\fill}}p{0.58\columnwidth}p{0.34\columnwidth}}
        \toprule
        Number of scenarios $Q$ & 1000 \\
        Simulation area $A$ (wrapped) & $\qty{2.5}{\kilo\meter} \times \qty{2.5}{\kilo\meter}$ \\
        Channel realizations $O$ & 100 \\
        UE density $\lambda_K$ & $\qty{40}{\per\square\kilo\meter}$ \\
        AP density $\lambda_L$ & $\qty{100}{\per\square\kilo\meter}$ \\
        CPU density $\lambda_J$ (distributed sites) & $\qty{10}{\per\square\kilo\meter}$ \\
        Antennas per AP $N$ & 4 \\
        Bandwidth & $\qty{20}{\mega\hertz}$ \\
        Rayleigh channel correlation $\mymat{R}$ & see \cite[Sec.~2.6]{björnsonMassiveMIMONetworks2017} \\
        Angular spread distribution & Gaussian, $\sigma_\varphi=15^\circ$ \\
        AP array geometry & Uniform linear array \\
        AP array orientation & Uniformly random \\
        AP antenna spacing & $0.5$ wavelengths \\
        Pathloss model & 3GPP Urban Micro \\
        Shadow fading model & 3GPP Urban Micro \\
        UE height & $\qty{1.5}{\meter}$ \\
        AP height & $\qty{10.0}{\meter}$ \\
        Carrier frequency $f_\mathrm{c}$ & $\qty{3.5}{\giga\hertz}$ \\
        Pilots $\tau_p$ & $10$ symbols \\
        Uplink/downlink data $\tau_u = \tau_d$ & $95$ symbols \\
        Coherence block $\tau_c = \tau_p+\tau_u+\tau_d$ & $200$ symbols \\
        Pilot power $\eta_k$ & $\qty{20}{\dBm}$ \\
        Uplink max power $P^\mathrm{max}_\mathrm{ul}$ & $\qty{20}{\dBm}$ \\
        Downlink max power $P^\mathrm{max}_\mathrm{dl}$ & $\qty{23}{\dBm}$ \\
        Measuring threshold $\beta^\mathrm{m}$ & $\qty{-140}{\dB}$ \\
        Estimation threshold $\beta^\mathrm{e}$ & $\qty{-120}{\dB}$ \\
        Candidate threshold $\beta^\mathrm{c}$ & $\qty{-120}{\dB}$ \\
        \bottomrule
    \end{tabular*}
\end{table}

\subsubsection{Simulation Area Size A}
Even with wrap-around, too small simulation areas introduce bias because periodic images of the network interact and distant interference is underrepresented.
In \autoref{tab:area_size}, we compare several side lengths $\sqrt{A}$ with a large-area reference simulation using $A = \qty{4.0}{\kilo\meter} \times \qty{4.0}{\kilo\meter}$.
For each area, we report the relative bias from the mean and 10\%-ile uplink \gls{se}, computed as the absolute difference to the reference divided by the reference value.
Small areas generally overestimate the \gls{se}, since they underestimate interference from distant \glspl{ue}.
The effect is stronger for 10\%-ile \glspl{ue}, which already have poor \gls{sinr}, and for cell-free systems, which cancel nearby interference well and are therefore more affected by weaker interferers.

For the commonly used $\sqrt{A} = \qty{1.0}{\kilo\meter}$, the 10\%-ile \gls{se} is overestimated by 22\% for centralized cell-free but only by 5\% for cellular, misleadingly enlarging their apparent gap.
With $\sqrt{A} \geq \qty{2.5}{\kilo\meter}$, the 10\%-ile bias is only a few percent for all systems while the simulation effort remains manageable.

For context, with the 3GPP Urban Microcell pathloss model at $f_\mathrm{c} = \qty{3.5}{\giga\hertz}$, the large-scale fading coefficient reaches our measuring threshold $\beta^\mathrm{m} = \qty{-140}{\dB}$ at roughly $\qty{1.0}{\kilo\meter}$.
Thus, even \glspl{ue} below this association threshold should not be neglected in the simulation.

\begin{table}
    \centering
    \small
    \caption{Relative bias of mean and 10\%-ile uplink \gls{se} per \gls{ue} compared to a reference area size of $A = \qty{4.0}{\kilo\meter} \times \qty{4.0}{\kilo\meter}$.}
    \label{tab:area_size}
    \input{table_files/grid_size_table.tex}
\end{table}

\subsubsection{Number of Channel Realizations O}
A sufficiently large number of channel realizations $O$ is required to accurately estimate the expectations in the \gls{se} expressions.
\autoref{tab:num_realizations} quantifies the variance and bias caused by too few channel realizations.
For each $O$, we fix one scenario $\mathcal{S}$ and recompute the \gls{se} of every \gls{ue} 1000 times using independent channel-realization sets.
The relative variance is the resulting variance divided by the mean \gls{se} of that \gls{ue}, averaged over all \glspl{ue}.
As a high-realization reference, we compute each \gls{ue}'s \gls{se} once using $O = 1000$ realizations.
The relative bias compares each \gls{ue}'s mean over the 1000 runs with this reference and then averages over \glspl{ue}.

The variance decreases approximately linearly with $O$, as expected for a sample-mean estimator.
For small $O$, the sample mean is too close to the drawn channels rather than the true expectation implied by our statistical-\gls{csi} model.
This underestimates the interference terms in the \gls{se} bounds that depend on the variance around the mean, and therefore overestimates the \gls{se}.
In other words, for too small $O$, the simulation no longer properly reflects our assumption that the detector has access only to statistical rather than instantaneous \gls{csi}.
For $O = 50$, the relative variance is already around $1\%$ for all systems; for $O = 100$, the remaining bias is acceptable.
We therefore use $O = 100$ in the remainder of this work.

\begin{table}
    \caption{Relative bias and variance of the uplink \gls{se} estimate per \gls{ue} for a fixed scenario compared to a reference number of realizations of $O=1000$.}
    \centering
    \small
    \input{table_files/num_realizations_table.tex}
    \label{tab:num_realizations}
\end{table}

%% file: table_files/grid_size_table.tex
\begin{tabular*}{\columnwidth}{@{\extracolsep{\fill}}llcccc@{}}
\toprule
System & $\sqrt{A}$ & $\SI{0.5}{\kilo\meter}$ & $\SI{1.0}{\kilo\meter}$ & $\SI{1.5}{\kilo\meter}$ & $\SI{2.5}{\kilo\meter}$ \\
\midrule
Cellular & Mean & 19.44\% & 4.24\% & 1.90\% & 0.32\% \\
 & 10\%-ile & 31.90\% & 5.34\% & 2.80\% & 1.01\% \\
\midrule
Distributed  & Mean & 31.06\% & 6.53\% & 2.74\% & 0.62\% \\
Cell-Free & 10\%-ile & 68.53\% & 11.53\% & 4.86\% & 1.15\% \\
\midrule
Centralized  & Mean & 44.28\% & 14.13\% & 5.85\% & 1.57\% \\
Cell-Free & 10\%-ile & 71.49\% & 22.37\% & 9.01\% & 2.77\% \\
\bottomrule
\end{tabular*}

%% file: table_files/num_realizations_table.tex
\begin{tabular*}{\columnwidth}{@{\extracolsep{\fill}}llcccc@{}}
\toprule
System & & $O=1$ & $O=10$ & $O=50$ & $O=100$ \\
\midrule
Cellular & Bias & 34.50\% & 5.65\% & 2.42\% & 2.09\% \\
& Var. & 33.35\% & 4.72\% & 1.17\% & 0.62\% \\
\midrule
Distributed  & Bias & 14.25\% & 2.89\% & 1.33\% & 1.22\% \\
Cell-Free& Var. & 18.00\% & 2.75\% & 0.65\% & 0.34\% \\
\midrule
Centralized  & Bias & 5.31\% & 0.66\% & 0.34\% & 0.33\% \\
Cell-Free& Var. & 4.71\% & 0.56\% & 0.12\% & 0.06\% \\
\bottomrule
\end{tabular*}

%% file: results.tex
\section{Numerical Evaluation} \label{sec:results}
We numerically compare the seven system types from \autoref{sec:properties}.
The parameter set from \autoref{tab:simulation_parameters} is used as a baseline, and we vary the main deployment parameters to assess how the architectures respond to different operating conditions.
\begin{figure*}[t]
    \makebox[\linewidth][r]{\input{tikz_images/legend.tikz}\hspace*{0.07cm}}
\end{figure*}
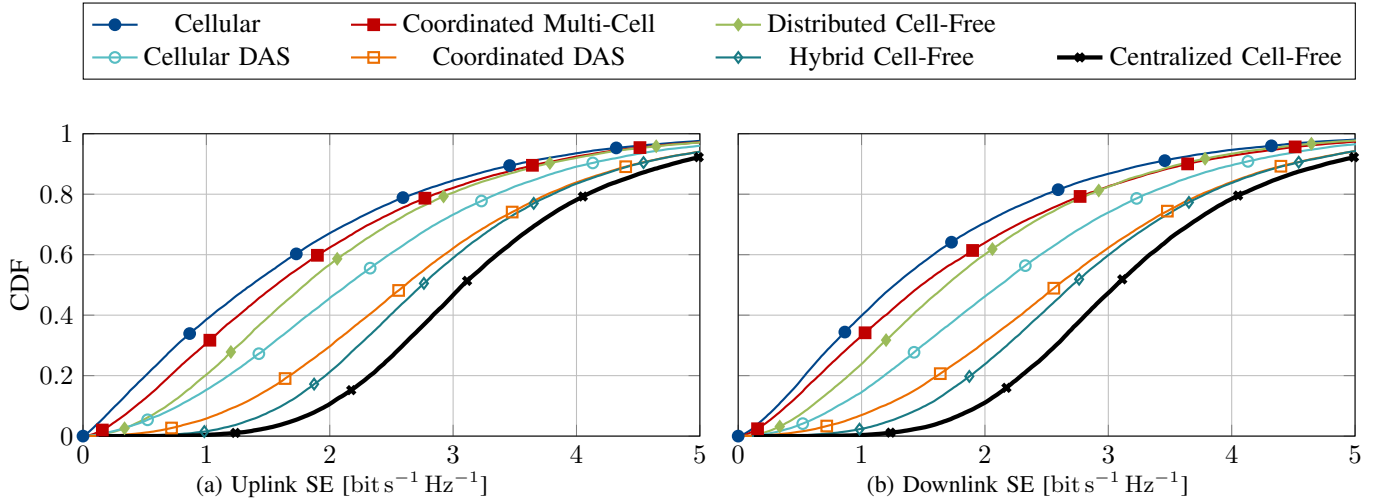
\begin{figure*}[t]
    \centering
    \captionsetup[subfigure]{skip=0pt}
    \begin{subfigure}{0.5\textwidth}
        \makebox[\linewidth][l]{\hspace*{-0.1cm}\input{tikz_images/example_performance_ul.tikz}}
        \caption{Uplink \gls{se} [\si{\bit\per\second\per\hertz}]}
    \end{subfigure}%
    \begin{subfigure}{0.5\textwidth}
        \makebox[\linewidth][l]{\hspace*{-0.5cm}\input{tikz_images/example_performance_dl.tikz}}
        \caption{Downlink \gls{se} [\si{\bit\per\second\per\hertz}]}
    \end{subfigure}
    \caption{\glspl{cdf} of the uplink and downlink \gls{se} per \gls{ue} under the baseline parameter set from \autoref{tab:simulation_parameters}.}
    \label{fig:example_performance}
\end{figure*}
\setlength{\fboxsep}{0pt}
\begin{figure*}[t]
    \centering
    \captionsetup[subfigure]{skip=0pt}
    \begin{subfigure}{0.5\textwidth}
        \makebox[\linewidth][l]{\hspace*{-0.1cm}\input{tikz_images/relative_capacity_ap_density.tikz}}
        \caption{\gls{ap} density $\lambda_L$ [\si{\per\square\kilo\meter}]} \label{subfig:relative_capacity_ap_density}
    \end{subfigure}%
    \begin{subfigure}{0.5\textwidth}
        \makebox[\linewidth][l]{\hspace*{-0.5cm}\input{tikz_images/relative_capacity_cpu_density.tikz}}
        \caption{\gls{cpu} density $\lambda_J$ [\si{\per\square\kilo\meter}]} \label{subfig:relative_capacity_cpu_density}
    \end{subfigure}

    \begin{subfigure}{0.5\textwidth}
        \makebox[\linewidth][l]{\hspace*{-0.1cm}\input{tikz_images/relative_capacity_antennas_per_ap.tikz}}
        \caption{Antennas per \gls{ap} $N$ (fixed $N\cdot\lambda_L = 400$)} \label{subfig:relative_capacity_antennas_per_ap}
    \end{subfigure}%
    \begin{subfigure}{0.5\textwidth}
        \makebox[\linewidth][l]{\hspace*{-0.5cm}\input{tikz_images/relative_capacity_ue_density.tikz}}
        \caption{\gls{ue} density $\lambda_K$ [\si{\per\square\kilo\meter}]} \label{subfig:relative_capacity_ue_density}
    \end{subfigure}
    \caption{Uplink 10\%-ile achieved relative capacity for varying deployment parameters.
    The baseline parameter set from \autoref{tab:simulation_parameters} is highlighted in each plot.}
    \label{fig:relative_capacity}
\end{figure*}
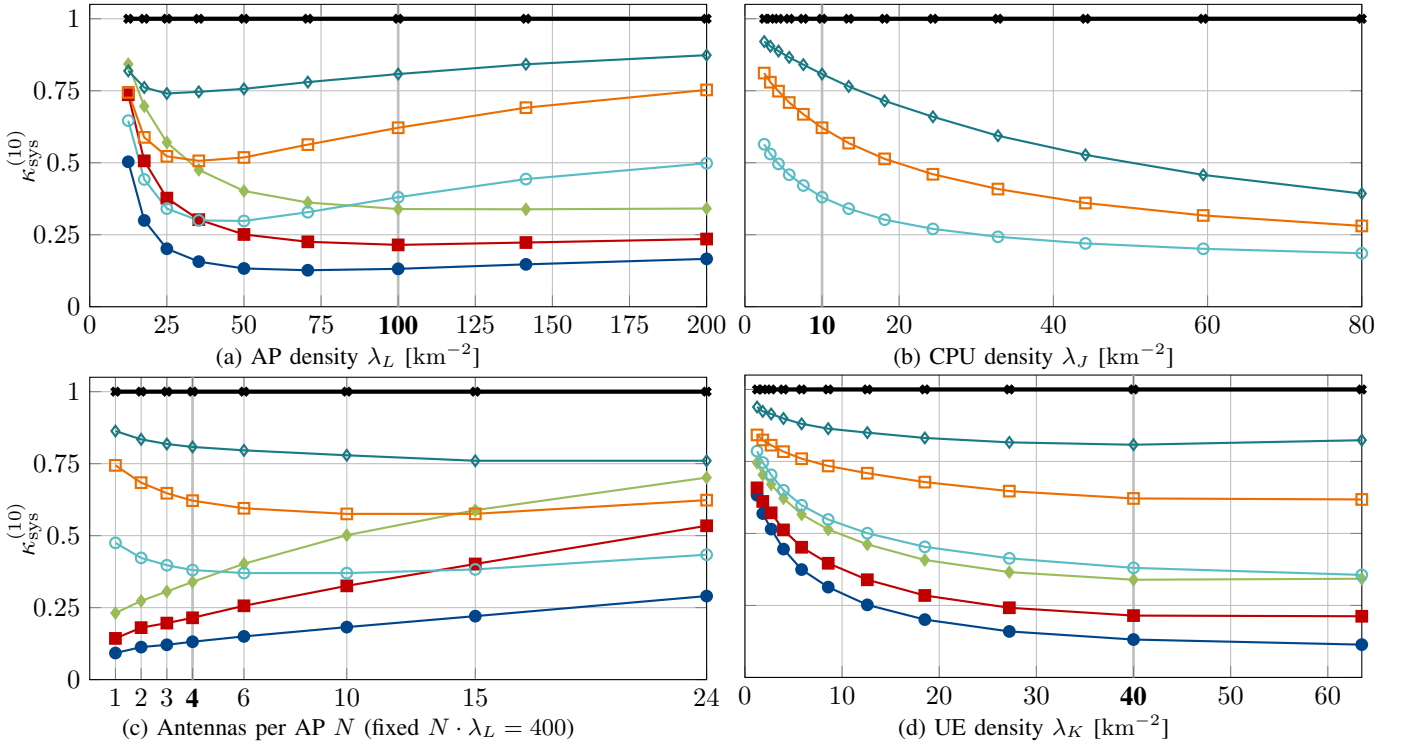
\subsection{Baseline Comparison of the Seven System Types} \label{subsec:baseline_comparison}
\autoref{fig:example_performance} shows the empirical \glspl{cdf} of the uplink and downlink \gls{se} per \gls{ue} of all seven system types.
The rightmost curve is the centralized cell-free system, an upper bound on all other systems due to its strongest cooperation and fully centralized processing.
We group the remaining six systems into those with \emph{colocated sites} (filled markers) and \emph{distributed sites} (empty markers).
Distributed-site systems generally outperform their colocated counterparts because they design combining and precoding vectors jointly across multiple \glspl{ap}, exploiting spatial diversity and mitigating interference more effectively than local per-\gls{ap} processing.

Within each group, the type of inter-site cooperation allowed via the midhaul links also strongly affects performance.
The purely cellular systems with \emph{no cooperation} perform worst.
Allowing \emph{user coordination} already improves performance: \glspl{ap} can estimate channels of \glspl{ue} in neighboring cells and include them in the sum of \autoref{eq:uplink_combiner}, enabling explicit interference cancellation.
All \glspl{ue} benefit from this interference cancellation, but the strongest \glspl{ue} improve little because they are already well served by their local \gls{ap} and suffer little inter-cell interference.
The gap between \emph{no cooperation} and \emph{user coordination} is especially large with distributed sites, where more combiner degrees of freedom make the added neighboring-\gls{ue} channel estimates more useful for interference cancellation.
\emph{User coordination} still leaves cell boundaries, so the weakest \glspl{ue} are limited not only by interference but also by low desired-signal power.
Allowing \emph{joint service} removes these cell boundaries and further improves these weak \glspl{ue}.
This effect is stronger with colocated sites, where cells are smaller and more \glspl{ue} lie at cell edges.

Downlink performance is qualitatively similar to uplink performance, indicating that the heuristic duality-based downlink processing is effective for each system type.
We therefore focus on the uplink in the following analysis, since similar conclusions also hold for the downlink.

\subsection{Achieved Relative Capacity} \label{subsec:cooperation_gain}
As explained in \autoref{sec:system_model}, the centralized cell-free system with the signal processing from \autoref{sec:processing} achieves the practically maximally achievable uplink ergodic \gls{se} per \gls{ue} under the assumptions of scalability, linear processing, pilot-based channel estimation, and equal power allocation across \glspl{ue}.
We therefore treat it as a natural upper bound, i.e., the \emph{practical capacity}.
We quantify how much of this upper bound each system achieves through the \emph{relative capacity at percentile $p$}, defined as
\begin{equation}
  \kappa_{\mathrm{sys}}^{(p)} = \frac{\mathrm{SE}_{\mathrm{sys}}^{(p)}}{\mathrm{SE}_{\mathrm{Centralized~CF}}^{(p)}}.
\end{equation}
\vspace{-0.2em}
The relative capacity lies between 0 and 1, where 1 means matching centralized cell-free performance at the \mbox{$p$-th} percentile of the \gls{se} per \gls{ue}.
Fairness is a central goal of cooperative architectures, so we focus on the weak \glspl{ue} and set $p=10$.
\autoref{fig:relative_capacity} shows that the achieved relative capacity depends on deployment parameters.
Again, we distinguish systems with \emph{colocated sites} (filled markers) from those with \emph{distributed sites} (empty markers).
Across all deployment parameters, within each group, \emph{limited joint service} achieves the highest relative capacity, followed by \emph{user coordination} and then \emph{no cooperation}, for the reasons outlined in \autoref{subsec:baseline_comparison}.

In \autoref{subfig:relative_capacity_ap_density}, we vary the \gls{ap} density $\lambda_L$.
For low \gls{ap} densities, cooperation provides little benefit: all systems are limited mainly by low desired-signal power rather than interference, so stronger interference cancellation yields small gains.
Good channels from multiple \glspl{ap} to the same \gls{ue} are also unlikely, so due to the thresholds in \autoref{alg:association_policy}, \glspl{ue} are often served by very few \glspl{ap} even in cell-free systems.
Thus, all system types achieve high relative capacity and small performance gaps at low \gls{ap} densities.
As $\lambda_L$ increases, the relative capacity of colocated-site systems first drops rapidly up to around $\lambda_L = 80 \, \mathrm{APs/km^2}$, then slightly increases again.
For distributed-site systems, the relative capacity decreases much less up to around $\lambda_L = 30 \, \mathrm{APs/km^2}$ and then quickly recovers.
A decrease in relative capacity does not imply worse absolute performance: all systems improve with denser \gls{ap} deployment, but centralized cell-free initially improves faster.
With distributed sites and fixed \gls{cpu} density, higher \gls{ap} density assigns more \glspl{ap} to each \gls{cpu}, giving the receive combiners more degrees of freedom for interference mitigation and spatial diversity.
With colocated sites, by contrast, the receive combiner dimension is always the number of antennas per \gls{ap}~$N$.
Thus, higher \gls{ap} density makes distributed sites increasingly favorable over colocated sites.

To better understand this effect, \autoref{subfig:relative_capacity_cpu_density} varies \gls{cpu} density $\lambda_J$ at fixed \gls{ap} density $\lambda_L$.
The achieved relative capacity of distributed-site systems gradually decreases with increasing \gls{cpu} density and converges to colocated-site performance for very high \gls{cpu} densities, where each \gls{cpu} is responsible for exactly one \gls{ap}.
The relationship among the three distributed-site systems also changes with \gls{cpu} density:
At low \gls{cpu} densities, \emph{coordinated \gls{das}} gains strongly over \emph{cellular \gls{das}} and approaches \emph{hybrid cell-free} performance.
At high \gls{cpu} densities, \emph{coordinated \gls{das}} performs only slightly better than \emph{cellular \gls{das}}, and both perform substantially worse than \emph{hybrid cell-free}.
The reason is that \emph{coordinated \gls{das}} gains over \emph{cellular \gls{das}} through better interference mitigation, while \emph{hybrid cell-free} gains over \emph{coordinated \gls{das}} by also increasing the desired-signal strength.
Low \gls{cpu} densities let many spatially distributed \glspl{ap} be jointly processed at one \gls{cpu}, which gives high flexibility for interference mitigation and therefore large \emph{coordinated \gls{das}} gains.
At high \gls{cpu} densities, interference mitigation is less effective, but increased desired-signal strength can partially compensate, making the \emph{hybrid cell-free} gain larger.

\autoref{subfig:relative_capacity_antennas_per_ap} further examines antenna distribution by varying the antennas per \gls{ap} $N$ while keeping the antenna density per area $N\cdot\lambda_L$ fixed, adapting $\lambda_L$ accordingly.
The more the antennas are separated across multiple \glspl{ap} (low $N$), the more important cooperation becomes.
In the extreme case of $N=1$, all colocated-site systems have low achieved relative capacity, while \emph{coordinated \gls{das}} and \emph{hybrid cell-free} approach centralized cell-free performance and \emph{cellular \gls{das}} remains substantially worse.
As $N$ increases, the achieved relative capacity of colocated-site systems improve because the local receive combiner dimension increases; at $N=15$, \emph{distributed cell-free} surpasses \emph{coordinated \gls{das}}.
Absolute performance is good for all systems at $N=1$, because the average distance between a \gls{ue} and its serving \gls{ap}(s) is small, but cooperation gains are much higher and distributed sites are especially beneficial.

Lastly, we vary the \gls{ue} density $\lambda_K$ in \autoref{subfig:relative_capacity_ue_density}.
Cooperation becomes more beneficial at higher \gls{ue} densities, where interference is the main limiting factor.
However, even for a moderately dense scenario of $\lambda_K = 20 \, \mathrm{UEs/km^2}$, the cellular system drops to an achieved relative capacity of $0.19$, showing that cooperation already provides substantial gains in this regime.

%% file: tikz_images/legend.tikz
\begin{tikzpicture}
    \begin{axis}[
            hide axis,
            xmin=0,
            xmax=1,
            ymin=0,
            ymax=1,
            width=\textwidth,
            height=1.8cm,
            every axis plot/.append style={thick},
            legend style={
                    draw=black,
                    fill=white,
                    legend columns=4,
                    /tikz/every even column/.append style={column sep=0.71cm},
                    at={(0.5,0.5)},
                    anchor=center,
                },
        ]
        \addlegendimage{color=tuhh_int_blue, mark=*, thick}
        \addlegendentry{Cellular}

        \addlegendimage{color=tuhh_int_red, mark=square*, thick}
        \addlegendentry{Coordinated Multi-Cell}

        \addlegendimage{color=tuhh_int_light_green, mark=diamond*, thick}
        \addlegendentry{Distributed Cell-Free}

        \addlegendimage{color=white, thick}
        \addlegendentry{ }

        \addlegendimage{color=tuhh_int_light_blue, mark=o, thick}
        \addlegendentry{Cellular DAS}

        \addlegendimage{color=tuhh_int_orange, mark=square, thick}
        \addlegendentry{Coordinated DAS}

        \addlegendimage{color=tuhh_int_green_blue, mark=diamond, thick}
        \addlegendentry{Hybrid Cell-Free}

        \addlegendimage{color=tuhh_int_black, mark=x, ultra thick}
        \addlegendentry{Centralized Cell-Free}
    \end{axis}
\end{tikzpicture}

%% file: tikz_images/example_performance_ul.tikz
\begin{tikzpicture}
    \begin{axis}[
            table/col sep=comma,
            xmin=0,
            xmax=5,
            ymin=0,
            ymax=1,
            extra y ticks={0},
            extra y tick labels={\phantom{0.00}},
            extra y tick style={
                grid=none,
                tick style={draw=none},
            },
            every axis plot/.append style={thick},
            ylabel={\phantom{$\kappa_{\mathrm{sys}}^{(10)}$}CDF\phantom{$\kappa_{\mathrm{sys}}^{(10)}$}},
            ylabel shift=-0.35cm,
            scale only axis,
            grid=major,
            height=4.0cm,
            width=0.90\linewidth,
        ]

        \addplot[color=tuhh_int_blue, mark=*, mark repeat=50, mark phase=0] table [x=x, y=ul]{plot_files/example_performance_cellular.csv};

        \addplot[color=tuhh_int_red, mark=square*, mark repeat=50, mark phase=10] table [x=x, y=ul]{plot_files/example_performance_coordinated_multi-cell.csv};
        
        \addplot[color=tuhh_int_light_green, mark=diamond*, mark repeat=50, mark phase=20] table [x=x, y=ul]{plot_files/example_performance_distributed_cf.csv};

        \addplot[color=tuhh_int_light_blue, mark=o, mark repeat=50, mark phase=30] table [x=x, y=ul]{plot_files/example_performance_cellular_das.csv};
        
        \addplot[color=tuhh_int_orange, mark=square,mark repeat=50, mark phase=40] table [x=x, y=ul]{plot_files/example_performance_coordinated_das.csv};

        \addplot[color=tuhh_int_green_blue, mark=diamond, mark repeat=50, mark phase=50] table [x=x, y=ul]{plot_files/example_performance_hybrid_cf.csv};

        \addplot[color=tuhh_int_black, style=ultra thick, mark=x, mark repeat=50, mark phase=60] table [x=x, y=ul]{plot_files/example_performance_centralized_cf.csv};
    \end{axis}
\end{tikzpicture}

%% file: tikz_images/example_performance_dl.tikz
\begin{tikzpicture}
    \begin{axis}[
            table/col sep=comma,
            xmin=0,
            xmax=5,
            ymin=0,
            ymax=1,
            extra y ticks={0},
            extra y tick labels={\phantom{0.00}},
            extra y tick style={
                grid=none,
                tick style={draw=none},
            },
            every axis plot/.append style={thick},
            ylabel={\phantom{\phantom{$\kappa_{\mathrm{sys}}^{(10)}$}CDF\phantom{$\kappa_{\mathrm{sys}}^{(10)}$}}},
            ylabel shift=-0.35cm,
            yticklabel={\phantom{\pgfmathprintnumber{\tick}}},
            scale only axis,
            grid=major,
            height=4.0cm,
            width=0.90\linewidth,
        ]

        \addplot[color=tuhh_int_blue, mark=*, mark repeat=50, mark phase=0] table [x=x, y=dl]{plot_files/example_performance_cellular.csv};
        \addplot[color=tuhh_int_red, mark=square*, mark repeat=50, mark phase=10] table [x=x, y=dl]{plot_files/example_performance_coordinated_multi-cell.csv};
        \addplot[color=tuhh_int_light_green, mark=diamond*, mark repeat=50, mark phase=20] table [x=x, y=dl]{plot_files/example_performance_distributed_cf.csv};
        \addplot[color=tuhh_int_light_blue, mark=o, mark repeat=50, mark phase=30] table [x=x, y=dl]{plot_files/example_performance_cellular_das.csv};
        \addplot[color=tuhh_int_orange, mark=square,mark repeat=50, mark phase=40] table [x=x, y=dl]{plot_files/example_performance_coordinated_das.csv};
        \addplot[color=tuhh_int_green_blue, mark=diamond, mark repeat=50, mark phase=50] table [x=x, y=dl]{plot_files/example_performance_hybrid_cf.csv};
        \addplot[color=tuhh_int_black, style=ultra thick, mark=x, mark repeat=50, mark phase=60] table [x=x, y=dl]{plot_files/example_performance_centralized_cf.csv};
    \end{axis}
\end{tikzpicture}

%% file: tikz_images/relative_capacity_ap_density.tikz
\begin{tikzpicture}
    \begin{axis}[
            table/col sep=comma,
            scale only axis,
            trim axis left,
            trim axis right,
            xmin=0,
            xmax=200,
            ymin=0,
            ymax=1.05,
            ytick={0,0.25,0.5,0.75,1.0},
            every axis plot/.append style={thick},
            ylabel={$\kappa_{\mathrm{sys}}^{(10)}$},
            ylabel shift=-0.35cm,
            grid=major,
            xtick={0,25,50,75,125,150,175,200},
            extra x ticks={100},
            extra x tick labels={100},
            extra x tick style={
                tick label style={font=\bfseries},
                grid=major,
                major grid style={line width=1.0pt},
            },
            height=4.0cm,
            width=0.90\linewidth,
        ]
        \addplot[color=tuhh_int_blue, mark=*, mark repeat=1, mark phase=0]
        table [x=lambda_l, y=cellular]{plot_files/relative_capacity_ap_density.csv};

        \addplot[color=tuhh_int_red, mark=square*, mark repeat=1, mark phase=0]
        table [x=lambda_l, y=coordinated_multi_cell]{plot_files/relative_capacity_ap_density.csv};

        \addplot[color=tuhh_int_light_green, mark=diamond*, mark repeat=1, mark phase=0]
        table [x=lambda_l, y=distributed_cf]{plot_files/relative_capacity_ap_density.csv};

        \addplot[color=tuhh_int_light_blue, mark=o, mark repeat=1, mark phase=0]
        table [x=lambda_l, y=cellular_das]{plot_files/relative_capacity_ap_density.csv};

        \addplot[color=tuhh_int_orange, mark=square, mark repeat=1, mark phase=0]
        table [x=lambda_l, y=coordinated_das]{plot_files/relative_capacity_ap_density.csv};

        \addplot[color=tuhh_int_green_blue, mark=diamond, mark repeat=1, mark phase=0]
        table [x=lambda_l, y=hybrid_cf]{plot_files/relative_capacity_ap_density.csv};

        \addplot[color=tuhh_int_black, style=ultra thick, mark=x, mark repeat=1, mark phase=0]
        table [x=lambda_l, y=centralized_cf]{plot_files/relative_capacity_ap_density.csv};
    \end{axis}
\end{tikzpicture}

%% file: tikz_images/relative_capacity_cpu_density.tikz
\begin{tikzpicture}
    \begin{axis}[
            table/col sep=comma,
            scale only axis,
            trim axis left,
            trim axis right,
            xmin=0,
            xmax=80,
            ymin=0,
            ymax=1.05,
            ytick={0,0.25,0.5,0.75,1.0},
            every axis plot/.append style={thick},
            ylabel={\phantom{$\kappa_{\mathrm{sys}}^{(10)}$}},
            ylabel shift=-0.35cm,
            grid=major,
            yticklabel={\phantom{\pgfmathprintnumber{\tick}}},
            xtick={0,20,40,60,80},
            extra x ticks={10},
            extra x tick labels={10},
            extra x tick style={
                tick label style={font=\bfseries},
                grid=major,
                major grid style={line width=1.0pt},
            },
            height=4.0cm,
            width=0.90\linewidth,
        ]



        \addplot[color=tuhh_int_light_blue, mark=o, mark repeat=1, mark phase=0]
        table [x=lambda_j, y=cellular_das]{plot_files/relative_capacity_cpu_density.csv};

        \addplot[color=tuhh_int_orange, mark=square, mark repeat=1, mark phase=0]
        table [x=lambda_j, y=coordinated_das]{plot_files/relative_capacity_cpu_density.csv};

        \addplot[color=tuhh_int_green_blue, mark=diamond, mark repeat=1, mark phase=0]
        table [x=lambda_j, y=hybrid_cf]{plot_files/relative_capacity_cpu_density.csv};

        \addplot[color=tuhh_int_black, style=ultra thick, mark=x, mark repeat=1, mark phase=0]
        table [x=lambda_j, y=centralized_cf]{plot_files/relative_capacity_cpu_density.csv};
    \end{axis}
\end{tikzpicture}

%% file: tikz_images/relative_capacity_antennas_per_ap.tikz
\begin{tikzpicture}
    \begin{axis}[
            table/col sep=comma,
            trim axis left,
            trim axis right,
            xmin=0,
            xmax=24,
            xtick={1,2,3,6,10,15,24},
            extra x ticks={4},
            extra x tick labels={4},
            extra x tick style={
                tick label style={font=\bfseries},
                grid=major,
                major grid style={line width=1.0pt},
            },
            ymin=0,
            ymax=1.05,
            ytick={0,0.25,0.5,0.75,1.0},
            every axis plot/.append style={thick},
            ylabel={$\kappa_{\mathrm{sys}}^{(10)}$},
            ylabel shift=-0.35cm,
            scale only axis,
            grid=major,
            height=4.0cm,
            width=0.90\linewidth,
        ]
        \addplot[color=tuhh_int_blue, mark=*, mark repeat=1, mark phase=0]
        table [x=N, y=cellular]{plot_files/relative_capacity_antennas_per_ap.csv};

        \addplot[color=tuhh_int_red, mark=square*, mark repeat=1, mark phase=0]
        table [x=N, y=coordinated_multi_cell]{plot_files/relative_capacity_antennas_per_ap.csv};

        \addplot[color=tuhh_int_light_green, mark=diamond*, mark repeat=1, mark phase=0]
        table [x=N, y=distributed_cf]{plot_files/relative_capacity_antennas_per_ap.csv};

        \addplot[color=tuhh_int_light_blue, mark=o, mark repeat=1, mark phase=0]
        table [x=N, y=cellular_das]{plot_files/relative_capacity_antennas_per_ap.csv};

        \addplot[color=tuhh_int_orange, mark=square, mark repeat=1, mark phase=0]
        table [x=N, y=coordinated_das]{plot_files/relative_capacity_antennas_per_ap.csv};

        \addplot[color=tuhh_int_green_blue, mark=diamond, mark repeat=1, mark phase=0]
        table [x=N, y=hybrid_cf]{plot_files/relative_capacity_antennas_per_ap.csv};

        \addplot[color=tuhh_int_black, style=ultra thick, mark=x, mark repeat=1, mark phase=0]
        table [x=N, y=centralized_cf]{plot_files/relative_capacity_antennas_per_ap.csv};
    \end{axis}
\end{tikzpicture}

%% file: tikz_images/relative_capacity_ue_density.tikz
\begin{tikzpicture}
    \begin{axis}[
            table/col sep=comma,
            scale only axis,
            trim axis left,
            trim axis right,
            xmin=0,
            xmax=63.5,
            ymin=0,
            ymax=1.05,
            ytick={0,0.25,0.5,0.75,1.0},
            every axis plot/.append style={thick},
            ylabel={\phantom{$\kappa_{\mathrm{sys}}^{(10)}$}},
            ylabel shift=-0.35cm,
            grid=major,
            yticklabel={\phantom{\pgfmathprintnumber{\tick}}},
            xtick={0,10,20,30,50,60},
            extra x ticks={40},
            extra x tick labels={40},
            extra x tick style={
                tick label style={font=\bfseries},
                grid=major,
                major grid style={line width=1.0pt},
            },
            height=4.0cm,
            width=0.90\linewidth,
        ]

        \addplot[color=tuhh_int_blue, mark=*, mark repeat=1, mark phase=0]
        table [x=lambda_k, y=cellular]{plot_files/relative_capacity_ue_density.csv};

        \addplot[color=tuhh_int_red, mark=square*, mark repeat=1, mark phase=0]
        table [x=lambda_k, y=coordinated_multi_cell]{plot_files/relative_capacity_ue_density.csv};

        \addplot[color=tuhh_int_light_green, mark=diamond*, mark repeat=1, mark phase=0]
        table [x=lambda_k, y=distributed_cf]{plot_files/relative_capacity_ue_density.csv};

        \addplot[color=tuhh_int_light_blue, mark=o, mark repeat=1, mark phase=0]
        table [x=lambda_k, y=cellular_das]{plot_files/relative_capacity_ue_density.csv};

        \addplot[color=tuhh_int_orange, mark=square, mark repeat=1, mark phase=0]
        table [x=lambda_k, y=coordinated_das]{plot_files/relative_capacity_ue_density.csv};

        \addplot[color=tuhh_int_green_blue, mark=diamond, mark repeat=1, mark phase=0]
        table [x=lambda_k, y=hybrid_cf]{plot_files/relative_capacity_ue_density.csv};

        \addplot[color=tuhh_int_black, style=ultra thick, mark=x, mark repeat=1, mark phase=0]
        table [x=lambda_k, y=centralized_cf]{plot_files/relative_capacity_ue_density.csv};
    \end{axis}
\end{tikzpicture}

%% file: conclusion.tex
\vspace{-1.1em}
\section{Conclusion} \label{sec:conclusion}
In this work we established a framework for a fair comparison of cellular and cooperative massive \gls{mimo} systems and used it to evaluate the performance of different system architectures with varying cooperation levels.
It was previously known that cell-free operation substantially outperforms small cell systems \cite{ngoCellFreeMassiveMIMO2017} and that centralized operation, although hard to practically implement, is necessary to fully exploit the potential of cell-free \gls{mimo} \cite{bjornsonMakingCellFreeMassive2020}.
Our work introduces more nuance into this picture by including more practical architectures into the comparison.
We identified two main concepts of cooperation, namely colocated vs. distributed sites and cooperation between sites, giving us a total of seven different system architectures to compare.
We evaluated the maximally achievable performance of each of these architectures for different deployment parameters and especially focused on the weak \glspl{ue}.
Importantly, to make this evaluation possible, we derived a new uplink \gls{se} bound, generalized practical and spectrally efficient signal processing to these architectures, and refined numerical Monte-Carlo methods to avoid biases.

Our results show that presence of distributed sites, i.e., joint beamforming across spatially distributed antennas, is the main driver of performance gains.
Hence, hybrid cell-free is an attractive option since it achieves most of the possible performance gains throughout all deployment parameters (\autoref{fig:relative_capacity}) while being practically feasible \cite{ranjbarCellFreeMMIMOSupport2022}.
For the envisioned regime of cell-free massive \gls{mimo} with dense \gls{ap} deployments ($\lambda_L \geq 100 \mathrm{APs/km^2}$) and few antennas per \gls{ap} ($N \approx 4$), the coordinated \gls{das} architecture also achieves very good performance.
On the other hand, distributed cell-free operation and \gls{comp}-like systems such as coordinated multi-cell miss out on delivering the promised gains of cooperation and generally only perform well in scenarios with sparse \gls{ap} deployments and many antennas per \gls{ap}.

Hybrid cell-free and coordinated \gls{das} therefore emerge as strong candidates for future mobile networks (see \cite{tsukamotoFullStackTestbedInterSite2025} for a practical testbed).
Compared with centralized cell-free operation, these architectures place weaker requirements on inter-site synchronization, payload exchange, and instantaneous \gls{csi} sharing, and may therefore be more robust to practical impairments, although a detailed robustness analysis remains an important direction for future work.